\documentstyle[psbox,aaspp4]{article} % for revise

\begin{document}

\title{MOLECULAR CLOUD FORMATION IN SHOCK-COMPRESSED LAYERS
\altaffilmark{1}}

\author{HIROSHI KOYAMA\altaffilmark{2} AND SHU-ICHIRO INUTSUKA}

\affil{Division of Theoretical Astrophysics, \\
National Astronomical Observatory, Mitaka, Tokyo 181-8588, Japan; \\
E-mail addresses: hkoyama@th.nao.ac.jp, inutsuka@th.nao.ac.jp }

\altaffiltext{1}{To appear in the April 10, 2000, Vol. 533 \#1 issue 
of the Astrophysical Journal.}
\altaffiltext{2}{ Also at The Department of Astronomy, 
School of Science, University of Tokyo,
Bunkyo-ku, Tokyo 113-0033, Japan}

%\accepted{November 17, 1999}

%\setlength{\baselineskip}{12pt}

\begin{abstract}
We investigate the propagation of a shock wave
into a warm neutral medium and cold neutral medium 
by one-dimensional hydrodynamic calculations
with detailed treatment of thermal and chemical processes.
Our main result shows that 
thermal instability inside the shock-compressed layer produces  
a geometrically thin, dense layer in which
a large amount of hydrogen molecules is formed.
Linear stability analysis suggests
that this thermally-collapsed layer will fragment
into small molecular cloudlets.
We expect that frequent compression due to supernova explosions,
stellar winds, spiral density waves, etc., in the galaxy 
make the interstellar medium occupied by these small molecular cloudlets.
\end{abstract}


\section{INTRODUCTION}
The tiny scale structures on scales of tens of astronomical units
were observed in the interstellar medium (ISM).
These structures have been detected by 21 cm absorption lines against 
quasars with VLBI techniques (Dieter, Welch, \& Romney 1976),
against pulsars with time variability (Frail et al. 1994) and
against close binary stars in optical interstellar lines
(Meyer \& Blades 1996).
The tiny scale structure is seen in all directions for which it has been 
searched, which suggests that it is ubiquitous,
not associated with large extinction (Heiles 1997).
AU-sized structures have been seen not only in atomic gas, but also
in molecular gas.
Langer et al. (1995) observed several clumps ranging in size from 
0.007 $-$ 0.021 pc 
in the Taurus molecular cloud 1 (TMC1) with interferometer techniques.
The mass of these smallest fragments is of the order of $0.01 M_{\sun}$.
These small-scale structures appear to be gravitationally unbound 
by a large factor.
The presence of these tiny scale atomic and molecular
structures in variety of the interstellar medium
suggests that their formation mechanisms have some similarities.

Models of ISM have been studied by many authors.
Field, Goldsmith, \& Habing (1969) first showed that 
cold ($T\sim 50$ K) neutral medium (CNM) and 
warm ($T\sim 8000$ K) neutral medium (WNM)
can be in pressure equilibrium.
Cox \& Smith (1974) showed that supernova explosions can greatly modify 
the general aspect of the ISM.
McKee \& Ostriker (1977) developed a theory of 
a supernova-dominated ISM.
The late phase supernova remnant (SNR) expands 
until the internal pressure drops to the
ambient interstellar pressure ($n$ = 0.1 cm$^{-3}$, \, $T$ = 10$^4$ K).
This maximum radius $R_{\rm max}$ = 10$^{2.1}$ pc is reached at a time
$t_{\rm max}$ = 10$^{6.3}$ yr.
The total volume $SR^3_{\rm max}t_{\rm max}$ which all SNRs occupy in the
maximum expansion time $t_{\rm max}$
exceeds the volume of the galactic disk,
where $S \sim$ 10$^{-2}$ yr$^{-1}$ is the supernova rate in our galaxy.
This means that the 
SNRs overlap before they are dissipated.
Thus, the study of shock propagation into ISM is important.

In this paper, we study the basic processes in the evolution of
ISM with a strong shock wave
by means of one-dimensional hydrodynamic calculations 
with non-equilibrium thermal and chemical processes. 
The processes we include in our calculation are 
photoelectric heating from small grains and PAHs, 
heating and ionization by cosmic rays and X-rays,
heating by H$_2$ formation and destruction,
atomic line cooling from hydrogen Ly$\alpha$, 
\ion{C}{2}, \ion{O}{1}, \ion{Fe}{2}, and \ion{Si}{2},
rovibrational line cooling from H$_2$ and CO, 
and atomic and molecular collisions with grains.
Effects of self-gravity and magnetic field are not treated in this paper.
In the next section, we explain the details of our numerical methods.
Results are presented in section 3.
In section 4, we discuss the formation of small molecular cloudlets.
Section 5 is for summary.

\section{NUMERICAL METHODS}
\subsection{Hydrodynamic Equations}

To study the thermal and dynamical evolution of shock-compressed layer,
we use a one-dimensional plane parallel numerical method.
The hydrodynamics module of our scheme is based on the second-order 
Godunov method. 
We use the result of Riemann problem at each cell interface to calculate 
numerical fluxes, 
without introducing any artificial viscosity,  
which characterizes the Godunov method.  
Higher-order Godunov method has been the ``state-of-the-art'' 
numerical method in last two decades (see e.g., van Leer 1997), 
and widely used in astrophysical context 
(see e.g., Truelove et al. 1998) .  
The hydrodynamic equations for the system are written as follows:
\begin{equation}
\frac{d\rho}{dt}+\rho\frac{\partial v}{\partial X}=0,
\end{equation}
\begin{equation}
\rho\frac{dv}{dt}=-\frac{\partial P}{\partial X},
\end{equation}
\begin{equation}
P=\left(1.1+x_{\rm e}-\frac{x_2}{2}\right)n k_{\rm B}T,
\end{equation}
\begin{equation}
\rho=1.4m_{{\rm H}}n,
\end{equation}
\begin{equation}
\frac{1}{\gamma-1}\frac{dP}{dt}
-\frac{\gamma}{\gamma-1}\frac{P}{\rho}\frac{d\rho}{dt}
=n\Gamma-n\Lambda
+\frac{\partial}{\partial X}K\frac{\partial T}{\partial X}.
\end{equation}
In the above equations, $n$ is the number density of hydrogen nuclei,
$\rho$ is the mass density of the gas,
$v$ is the velocity of fluid elements, $P$ is the gas pressure,
and $K$ is the coefficient of thermal conductivity.
We have assumed an abundance of He atom is $0.1 n$.
For the range of temperatures and ionized fractions considered,
the dominant contribution to thermal conductivity is that 
due to neutral atoms, for which
$K = 2.5 \times 10^3 T^{1/2}\,{\rm cm^{-1} K^{-1} s^{-1}}$ (Parker 1953).
We use the ratio of the specific heats $\gamma=5/3$ for simplicity. 
$\Gamma$ and $\Lambda$ are heating and cooling rate per hydrogen nucleus,
respectively.
These rates depend on number density $n$, temperature $T$, 
electron fraction $x_{\rm e}=n({\rm e^-})/n$, 
H$_2$ fraction $x_2=2n({{\rm H}_2})/n$, 
and CO fraction $x_{{\rm CO}}=n({{\rm CO}})/n$.
The gas behind the shock front is chemical non-equilibrium state.
Therefore we solve three chemical equations for 
$x_{\rm e}$, $x_2$, and $x_{{\rm CO}}$ 
and hydrodynamic equations simultaneously.

\subsection{Heating and Cooling Processes}
\begin{figure}[htbp]
\psbox[xsize=0.49\hsize]{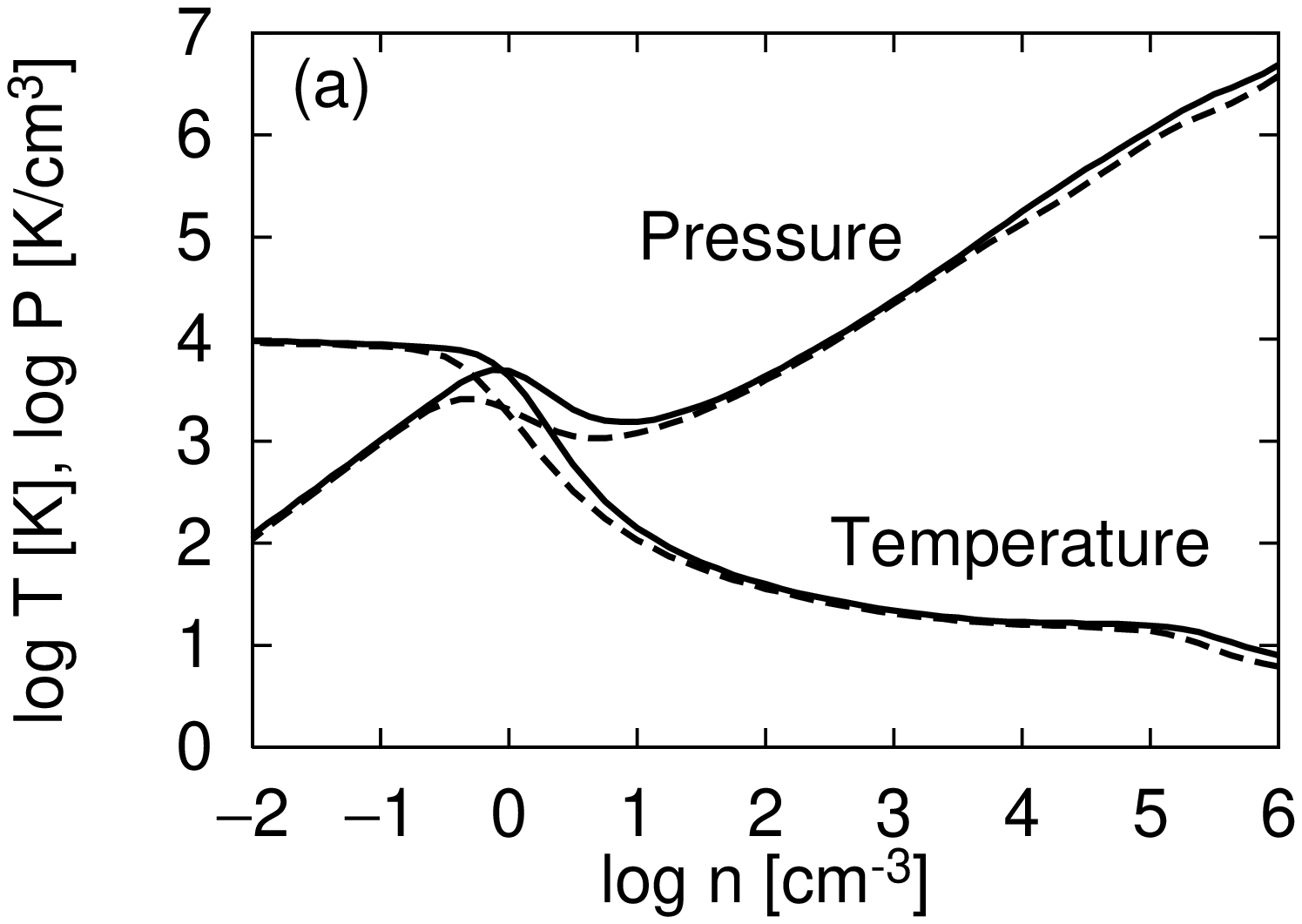}
\psbox[xsize=0.49\hsize]{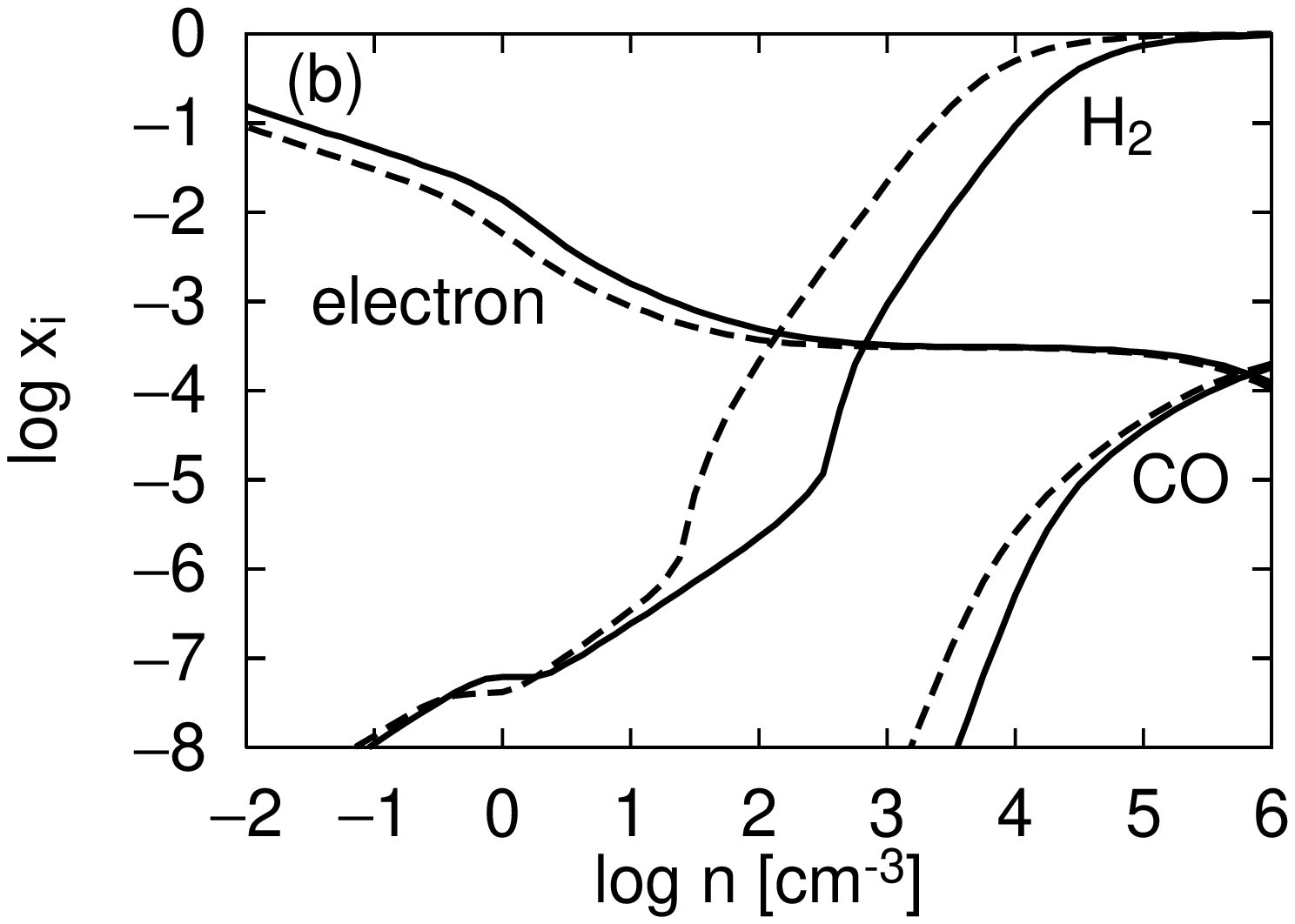}

\psbox[xsize=0.49\hsize]{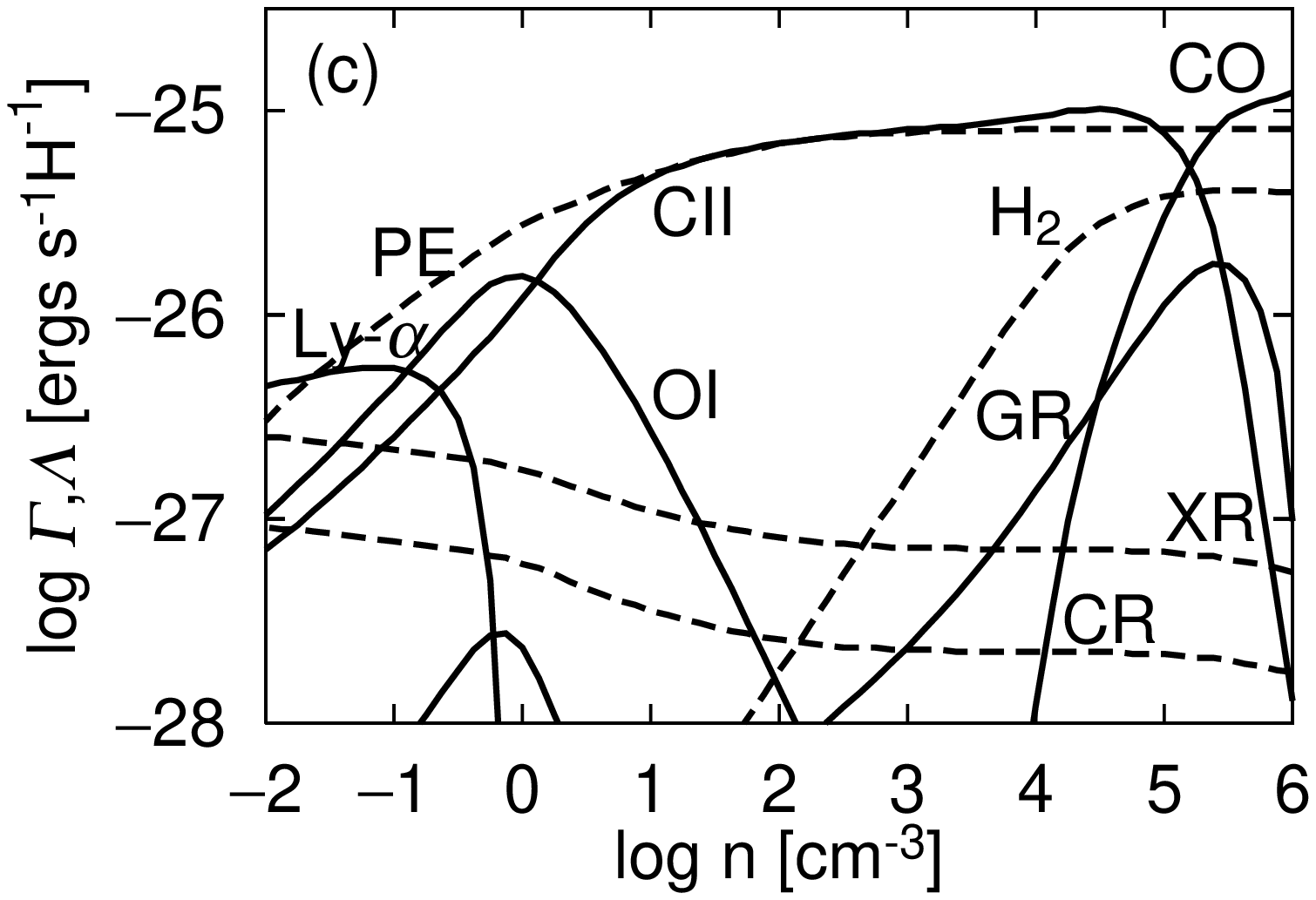}
\psbox[xsize=0.49\hsize]{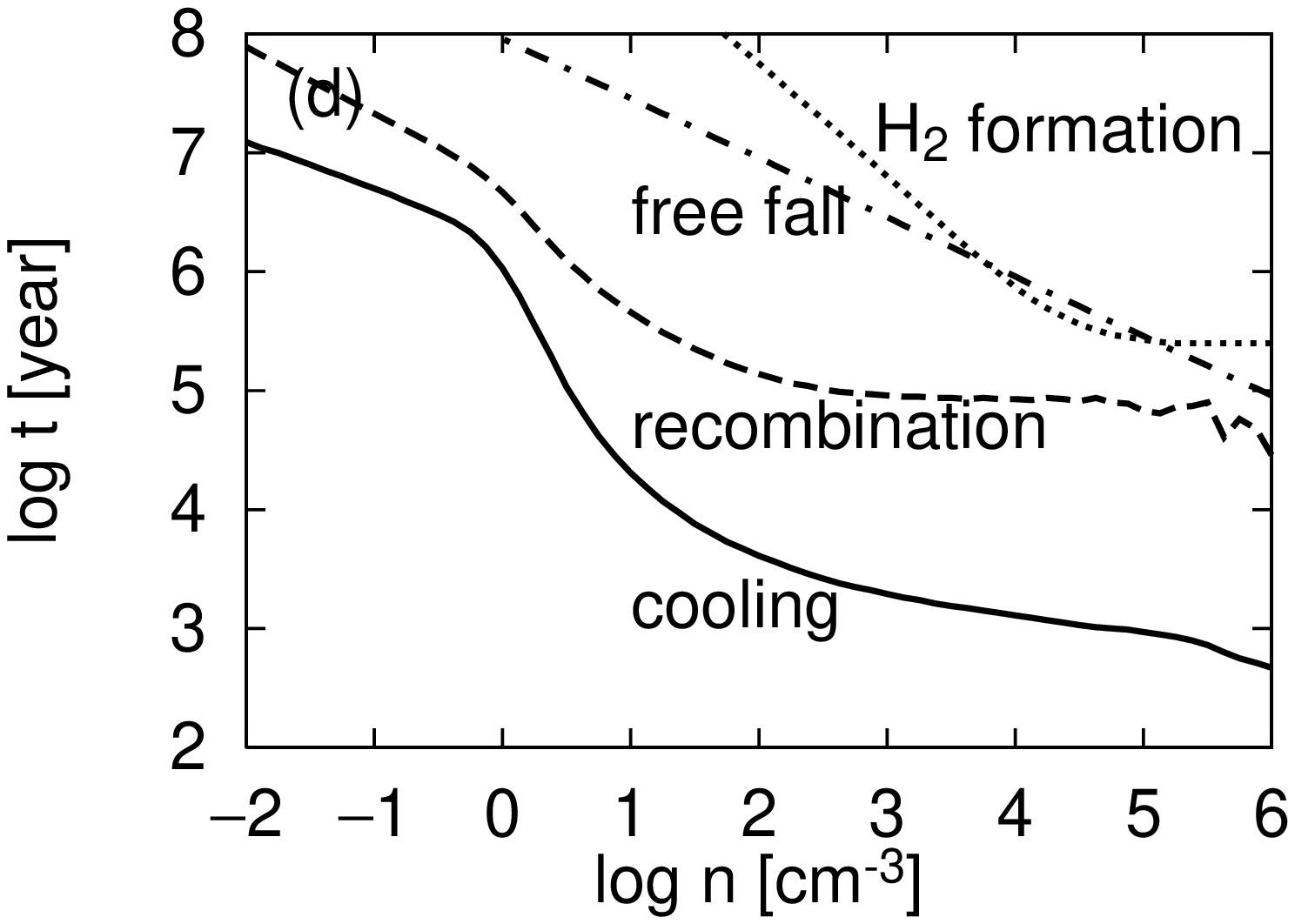}
\caption{(a) Equilibrium temperature and pressure. 
An absorbing column is 10$^{19}\, {\rm cm^{-2}}$ (solid lines),
10$^{20}\, {\rm cm^{-2}}$ (dash lines). 
(b) Chemical fractions as functions of number density. 
(c) Heating (dashed lines) and cooling rates (solid lines) per hydrogen
 nucleus in equilibrium condition corresponding to panel a (solid lines).
Heating processes are,
photoelectric heating from small grains and PAHs (PE),
X-ray (XR),
Cosmic-ray (CR), and
H$_2$ formation/destruction heating (H$_2$).
Cooling processes are
CII fine-structure (CII),
OI fine-structure (OI), 
Hydrogen Lyman-$\alpha$ (Ly-$\alpha$),
CO rotation/vibration line (CO), and
Atomic and molecular collisions with dust grains (GR).
(d) Comparison of various timescales,
cooling time (solid line), 
recombination time (dashed line), 
free fall time (dot-dash line),
and H$_2$ formation time (dotted line) 
which is defined by $(Rn_{\rm HI})^{-1}$.} 
\label{fig:equilibrium}
\end{figure}

Wolfire et al. (1995, hereafter W95) calculated the thermal equilibrium state 
in the neutral atomic phase.
We follow their calculation for $n<10^3 \, {\rm cm^{-3}}$.
We also calculate molecular processes for 
$10^3 \, {\rm cm^{-3}}<n<10^6 \, {\rm cm^{-3}}$
(see Figure \ref{fig:equilibrium}).
Contributions to the heating of an interstellar cloud are from 
the photoelectric emission from small grains and PAHs, 
ionization by cosmic rays and soft X-rays,
and the formation and photodissociation of H$_2$.
Following W95, the local FUV field is set to 1.7 times Habing's estimate 
(i.e., $G_0=1.7$).
The cooling function is dominated by line emission from H, 
C, O, Si, and Fe, 
by rovibrational lines from H$_2$ and CO, 
and by atomic and molecular collisions with dust grains.
To solve these thermal processes in non-equilibrium, 
we solve a set of 
three
time-dependent equations for ionization and recombination 
of hydrogen, and formation and dissociation of H$_2$ and CO.
Self-shielding effects are also included in the calculation of 
H$_2$ photodissociation.
The details of these processes are presented in Appendix \ref{THERMAL}.  
Figure \ref{fig:equilibrium} shows the equilibrium temperature, pressure,
and chemical fractions as functions of number density.
Important physical timescales are also shown in Figure \ref{fig:equilibrium}.

\subsection{Initial and Boundary Conditions}

\begin{figure}[htbp]
\psbox[xsize=0.49\hsize]{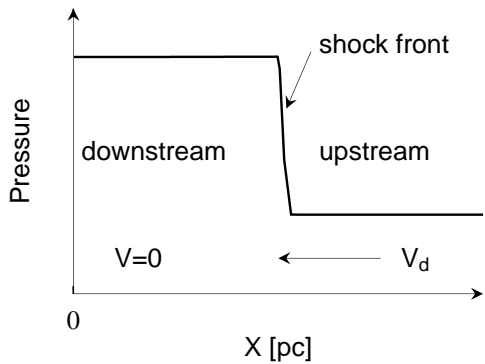}
\caption{Schematic configuration of our calculation.
The shock front propagates from left to right.
The velocity difference, $V_{\rm d}$, is defined by 
upstream velocity minus downstream velocity which we set to be zero.}
\label{fig:configure}
\end{figure}

The strong shock wave is characterized by Mach number. 
To analyze the evolution of ISM swept up by a strong shock wave,
we set up a plane-parallel cloud which collides against a rigid wall.
Figure \ref{fig:configure} shows the schematic configuration of
our calculation.
This configuration also corresponds to a face-on collision between 
two identical clouds.

We have generated a grid of calculations for
varying initial densities and collision velocities.
The typical models are summarized in table \ref{tbl-1}. 
Initial temperatures and chemical fractions are taken to be
the value of thermal and chemical equilibrium state
(see Figure \ref{fig:equilibrium}).
The velocity difference, 
$V_{\rm d}$, defined by upstream velocity minus downstream 
velocity corresponds to the range of Mach number $M = 2 - 12$.
This range of Mach number in the WNM ($T \sim 8000$ K) corresponds to
the sweeping-up speed of momentum-driven SNRs.
External FUV and X-ray radiation penetrate from both sides of  
two identical clouds.
The extinction of the radiation is evaluated by 
the cumulative column density of each grid.
The largest total column density we considered is that of the
standard \ion{H}{1} cloud ($10^{19 - 20}\,{\rm cm^{-2}}$).

\section{RESULTS}\label{section:RESULTS}

\subsection{Shock Propagation into WNM}\label{subsection:WNM}

\begin{figure}[htbp]
\psbox[xsize=0.49\hsize]{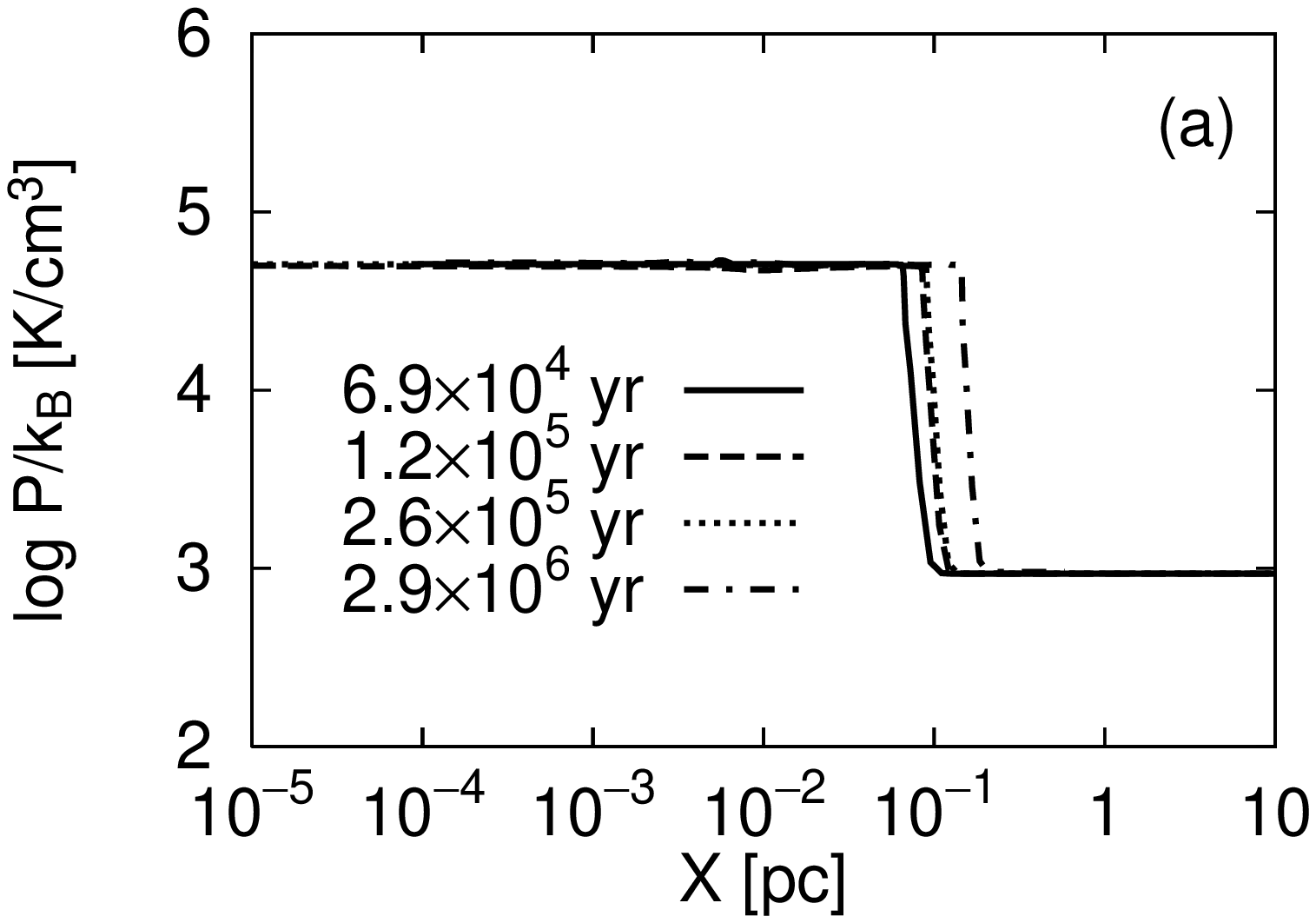}
\psbox[xsize=0.49\hsize]{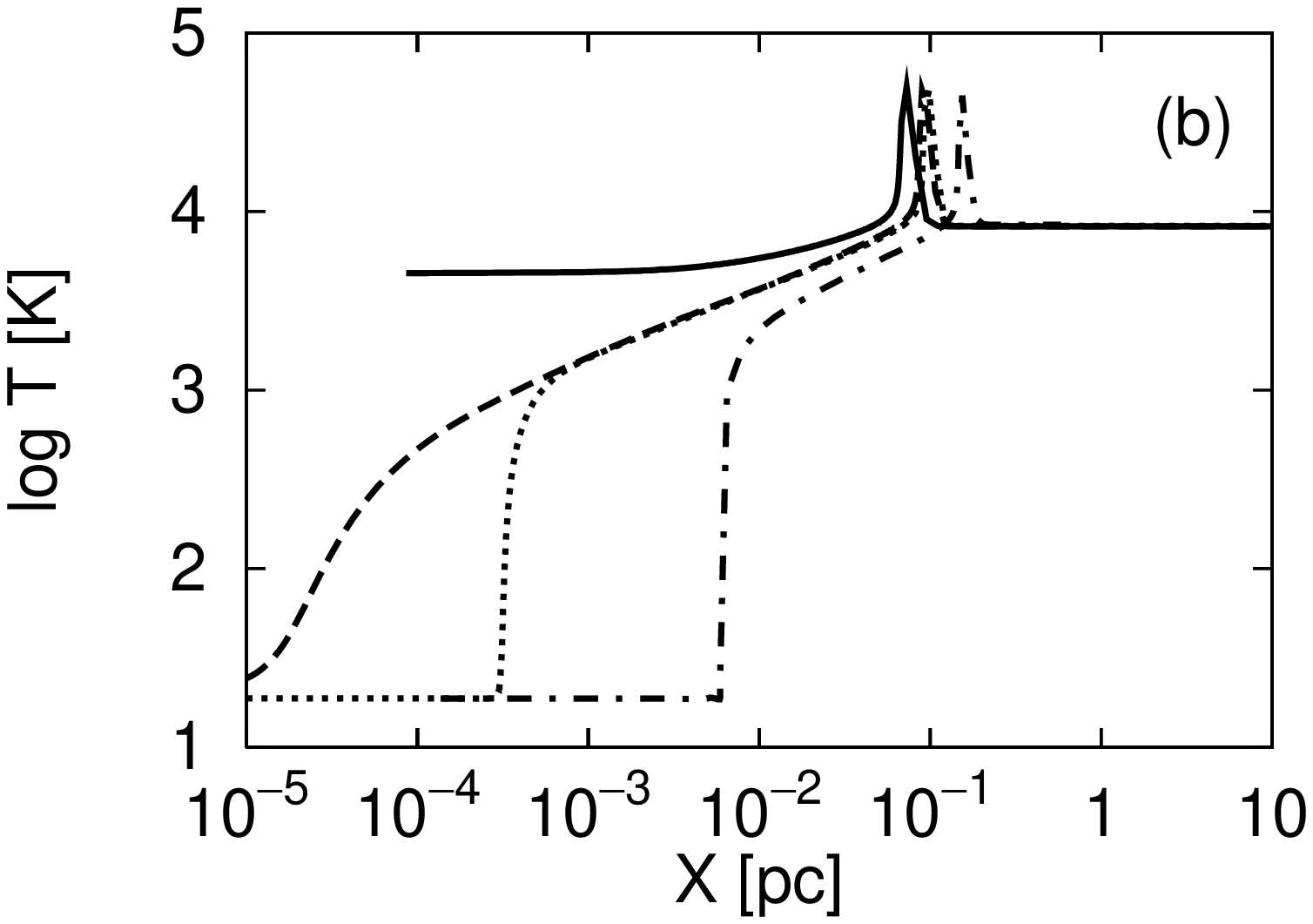}

\psbox[xsize=0.49\hsize]{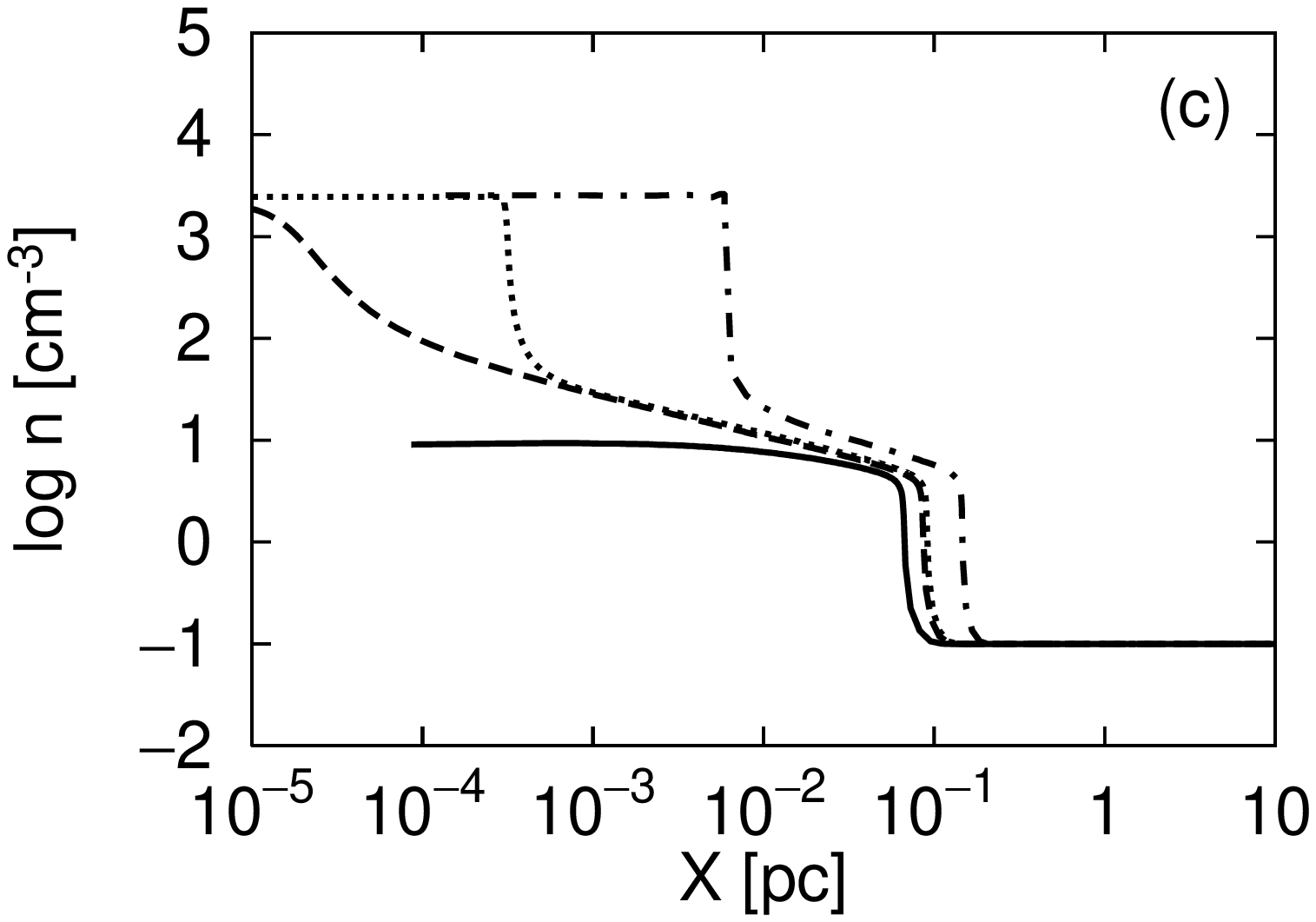}
\psbox[xsize=0.49\hsize]{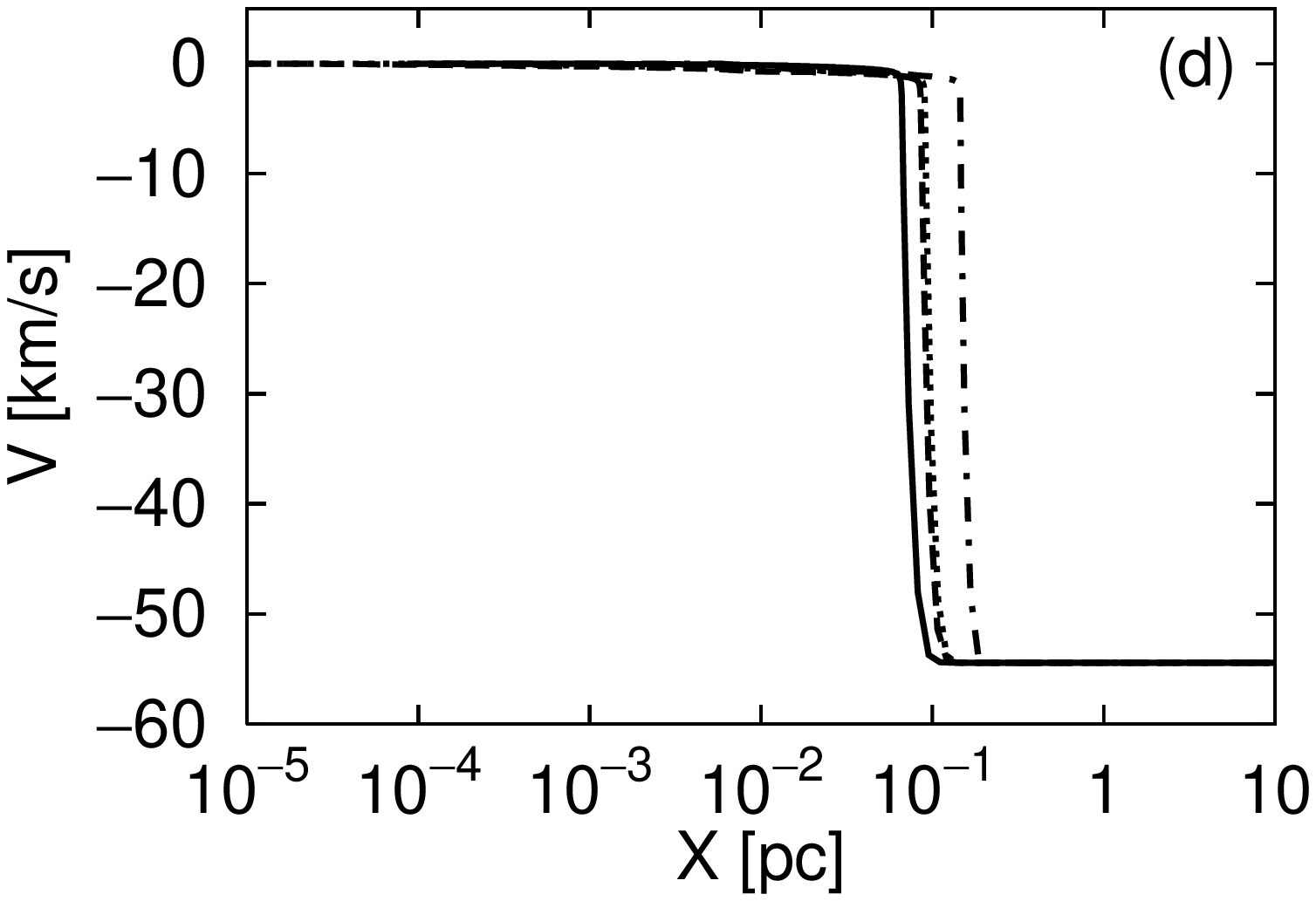}

\psbox[xsize=0.49\hsize]{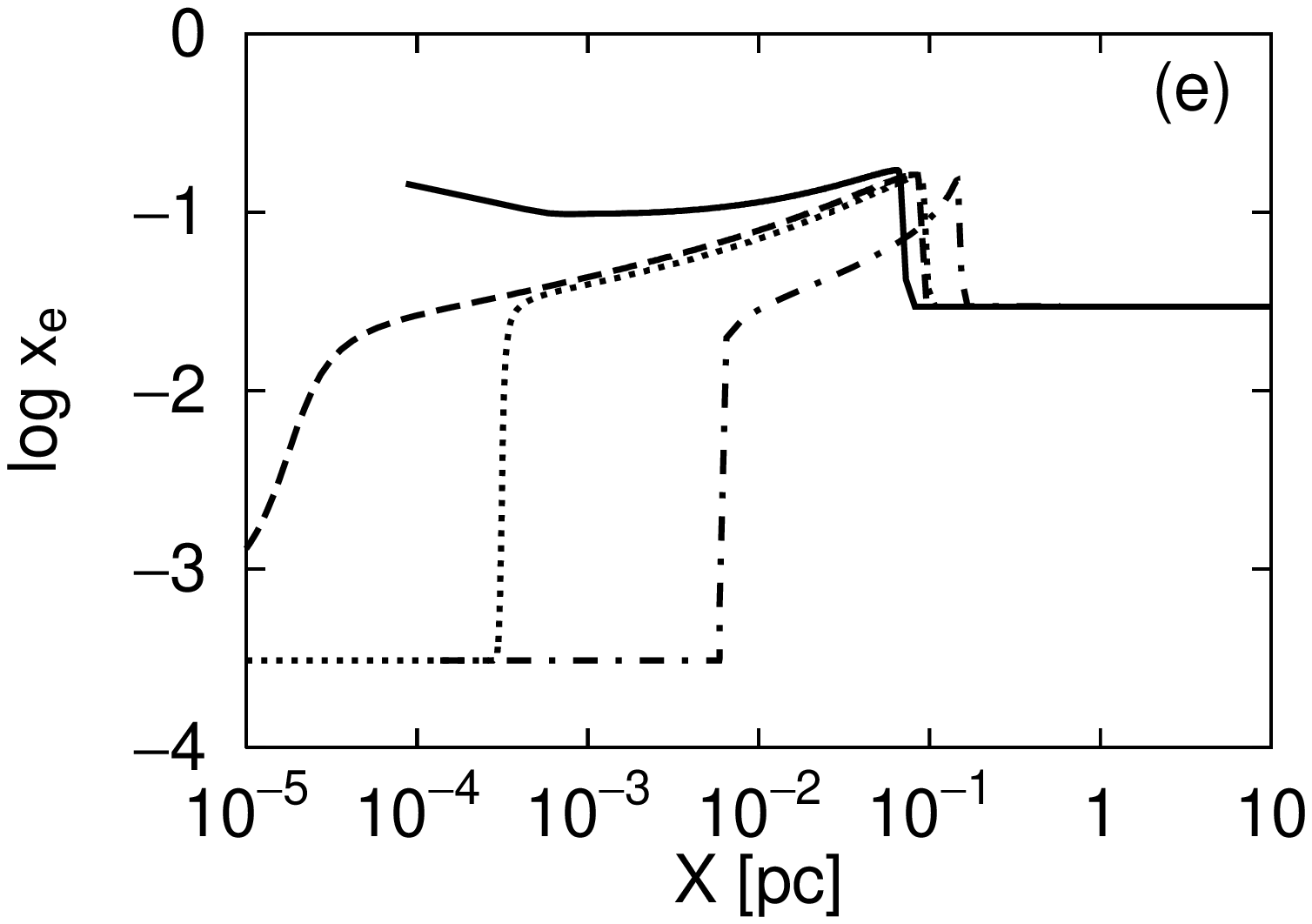}
\psbox[xsize=0.49\hsize]{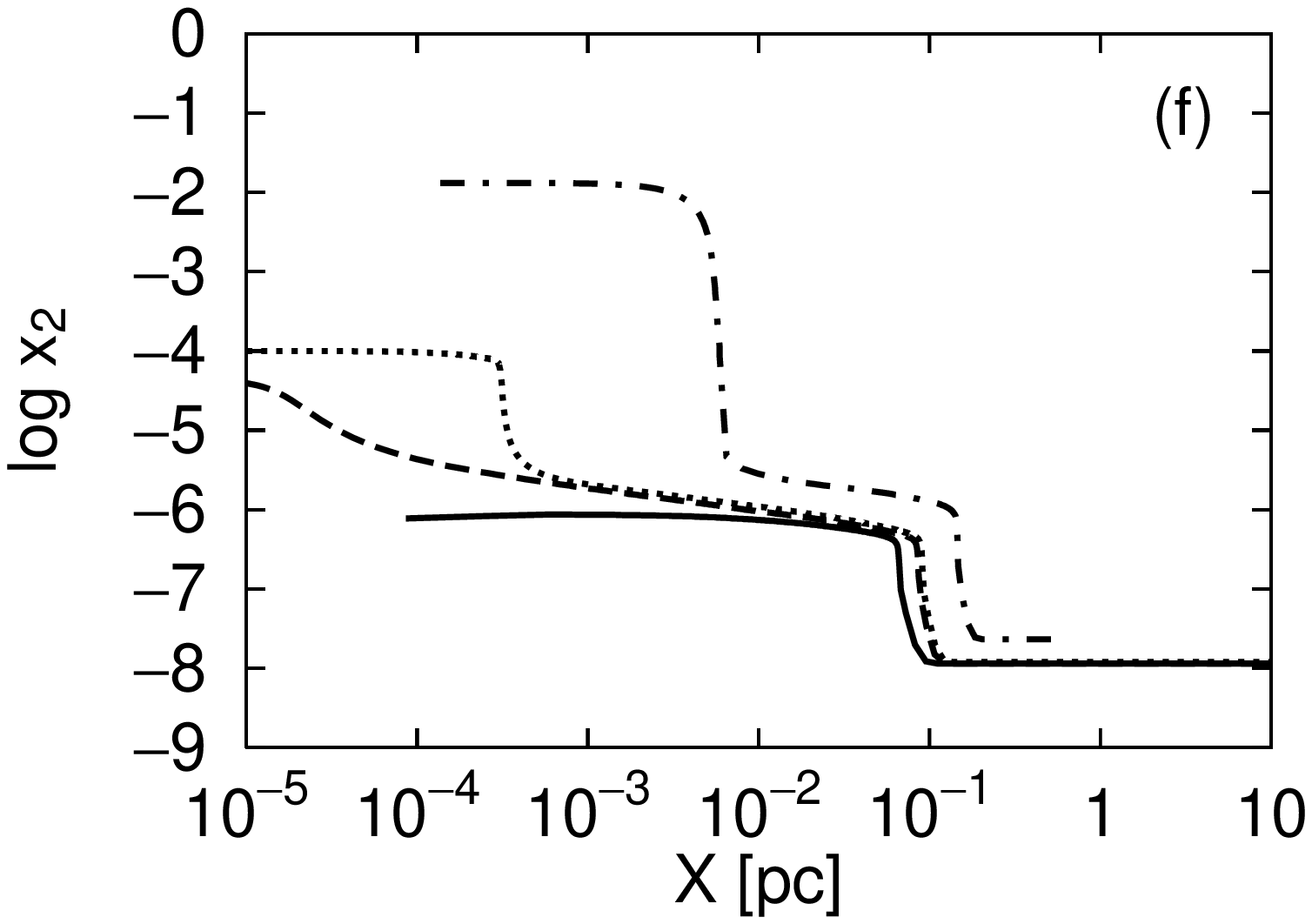}
\caption{The evolution of the shock propagation into the WNM.
The figures show (a) pressure, (b) temperature,
(c) number density of hydrogen nuclei,
(d) velocity, (e) electron number fraction,
and (f) H$_2$ number fraction.
Solid lines denote the compressed layer at $t=8.0 \times 10^4$ yr,
dashed lines at $t=1.3 \times 10^5$ yr,
dotted lines at $t=2.5 \times 10^5$ yr, 
and dot-dash lines at $t=2.9 \times 10^6$ yr.}
\label{fig:WNM Layer}
\end{figure}

In this section, we analyze the model W6 where a shock wave propagates
into WNM.
Figure \ref{fig:WNM Layer} shows the result of calculation.
Snapshots at 
$t=8.0\times 10^4$, $1.3\times 10^5$, $2.5\times 10^5$, and $2.9\times 10^6$ 
yr are shown.
The horizontal axis in logarithmic scale denotes distance from the rigid wall.
Figure \ref{fig:WNM Layer}a shows the evolution of pressure.
The shock-compressed layer is almost isobaric.
The shock front propagates from left to right.
Figure \ref{fig:WNM Layer}b shows the evolution of temperature.
Lyman $\alpha$ cooling is so efficient that the temperature of
post-shock gas quickly decreases to the pre-shock value.
In the next subsection, we analyze the shock front in detail.
Behind the shock front, cooling dominates heating because density is larger
and temperature decreases monotonically.
Around $t \simeq 10^5$ yr, 
temperature starts to decline rapidly because of thermal instability.
This instability is mainly driven by \ion{O}{1} (63 \micron) and \ion{C}{2} 
(157 \micron) line cooling. 
Eventually temperature attains thermal equilibrium value 
$T \simeq 19$ K at $n \simeq 2.5\times 10^3 \, {\rm cm^{-3}}$. 
Main coolant of collapsed layer is \ion{C}{2} (see 
Figure \ref{fig:equilibrium}c). 
Figure \ref{fig:WNM Layer}c shows the evolution of number density.
A thin layer is formed 
by thermal instability in shock-compressed layer.
The width of thermally-collapsed layer is much shorter than the
shock-compressed layer.
However column density of the thermally-collapsed layer becomes
comparable after $t=2.6\times 10^5$ yr (dotted line on Figure
\ref{fig:WNM Layer}c).
Figure \ref{fig:WNM Layer}d shows the evolution of velocity.
Figure \ref{fig:WNM Layer}e shows the evolution of electron fraction.
High shock temperature raises the ionization degree in the post-shock gas.
The timescale of recombination of electron in post-shock layer 
is larger than the timescale of gas cooling. 
Figure \ref{fig:WNM Layer}f shows 
the evolution of H$_2$ number fraction.
Because the timescale of H$_2$ formation is much
larger than cooling timescale (see Figure \ref{fig:equilibrium}d),
this thermal instability is not driven by H$_2$ cooling.
In the highest density region at final time step (dot-dash),
1.3\% of the hydrogen is in H$_2$ and 
0.0002\% of the carbon is in CO.
Note that the 
pre-shock H$_2$ number fraction at final time step (dot-dash) is larger
than the initial fraction,
because H$_2$ photodissociation radiation from the left hand
side is shielded by large H$_2$
column density in the collapsed layer in this one-dimensional
plane-parallel model.

\subsection{Structure Across The Shock Front}

\begin{figure}[htbp]
\psbox[xsize=0.49\hsize]{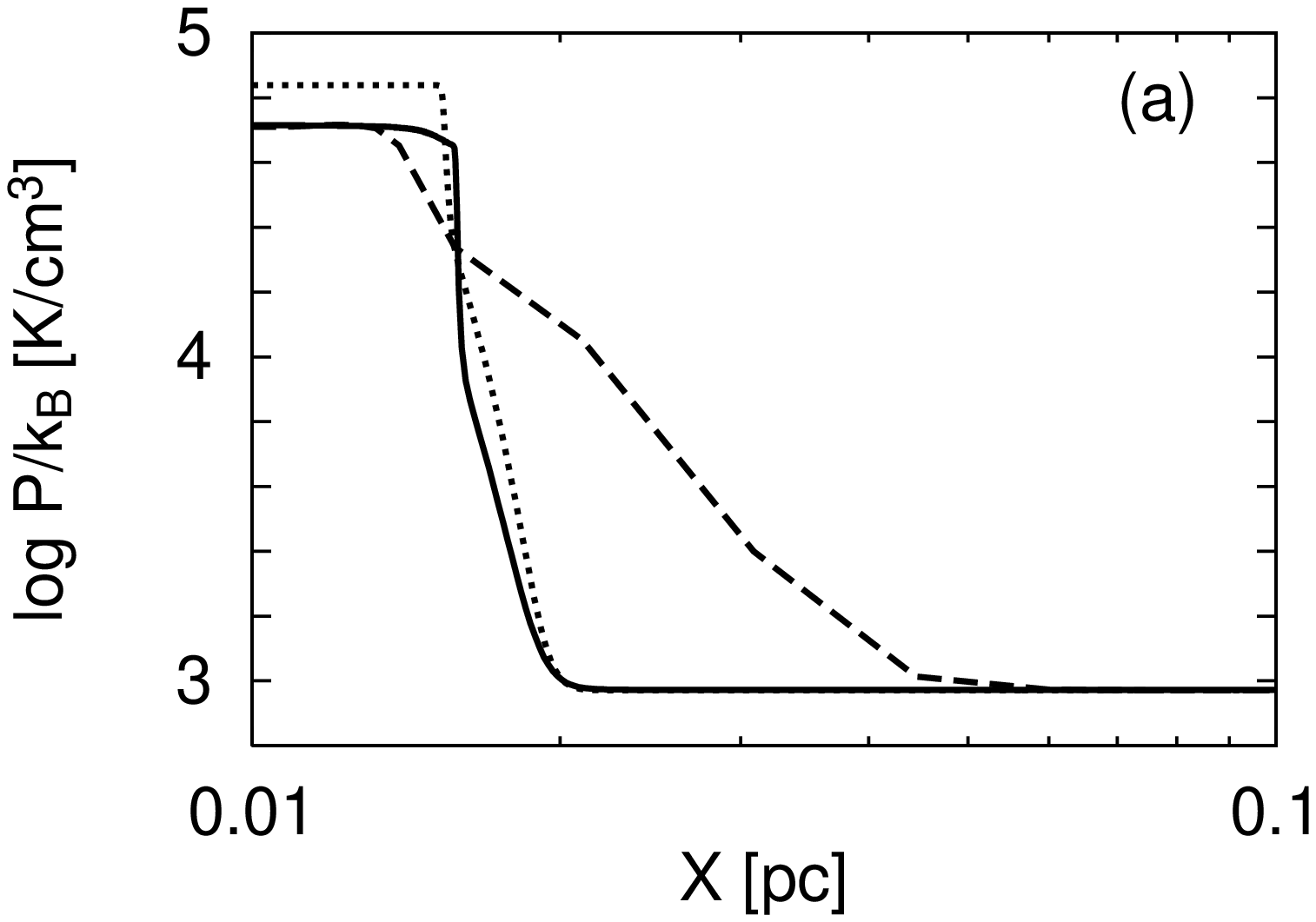}
\psbox[xsize=0.49\hsize]{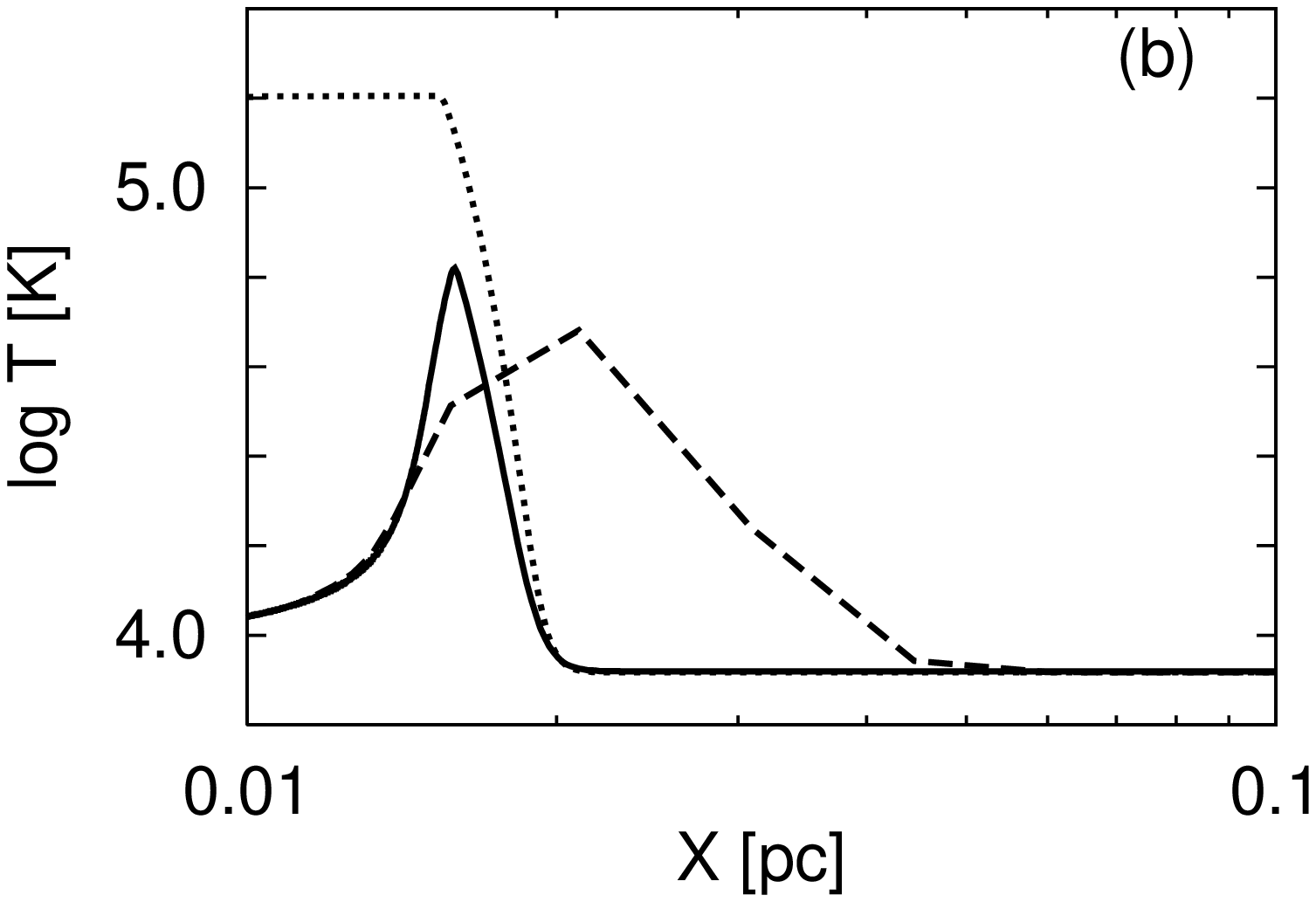}

\psbox[xsize=0.49\hsize]{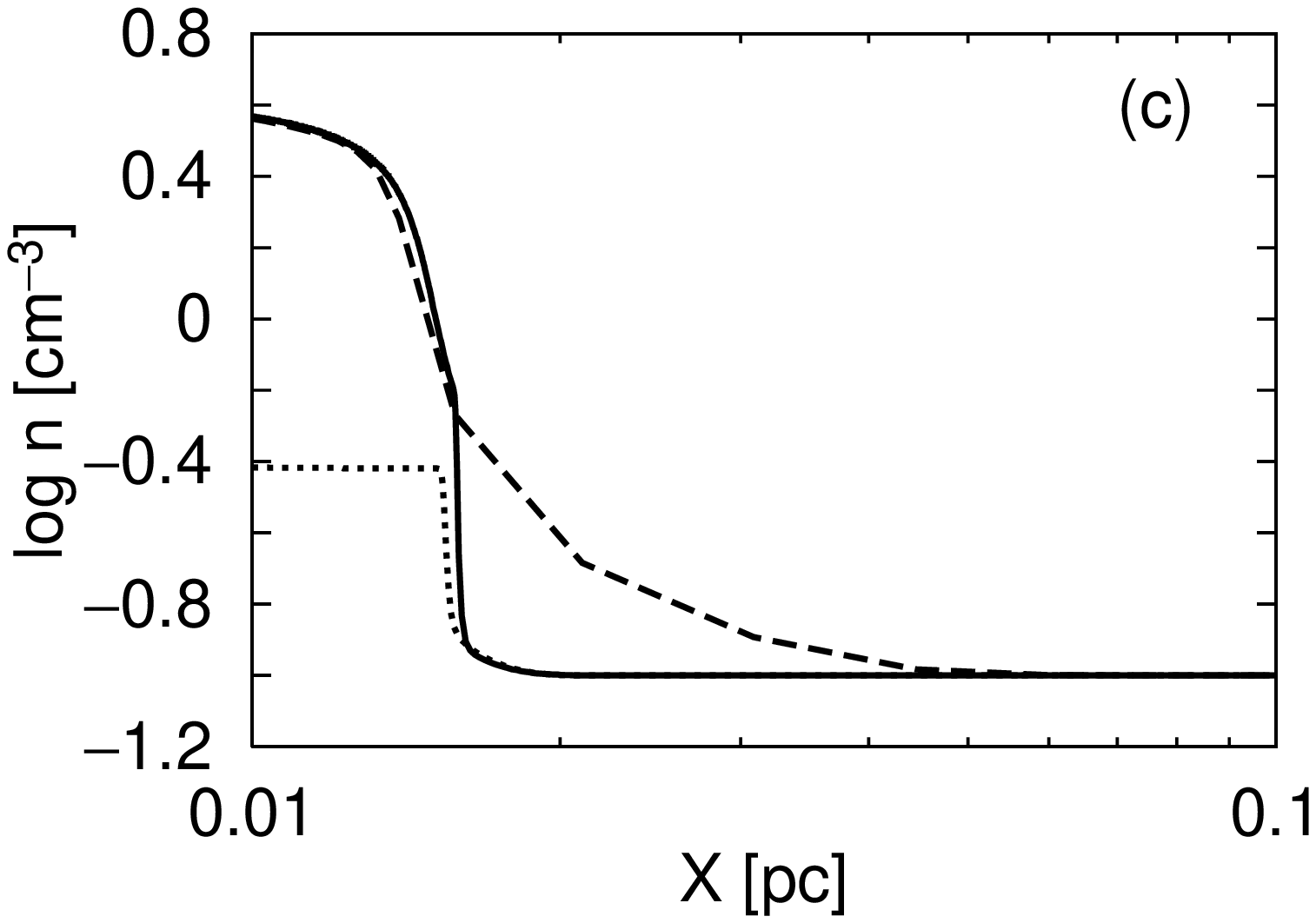}
\psbox[xsize=0.49\hsize]{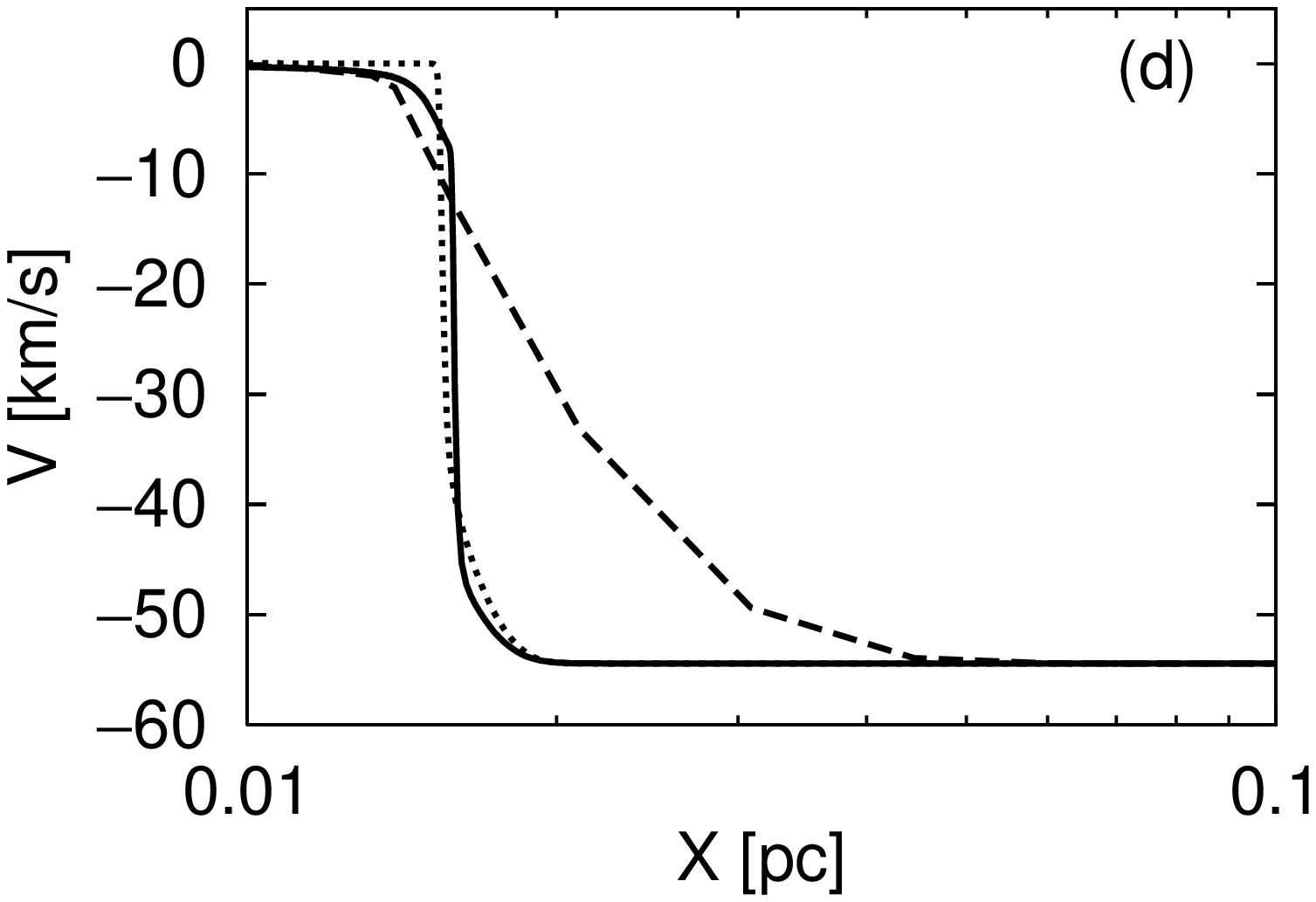}

\psbox[xsize=0.49\hsize]{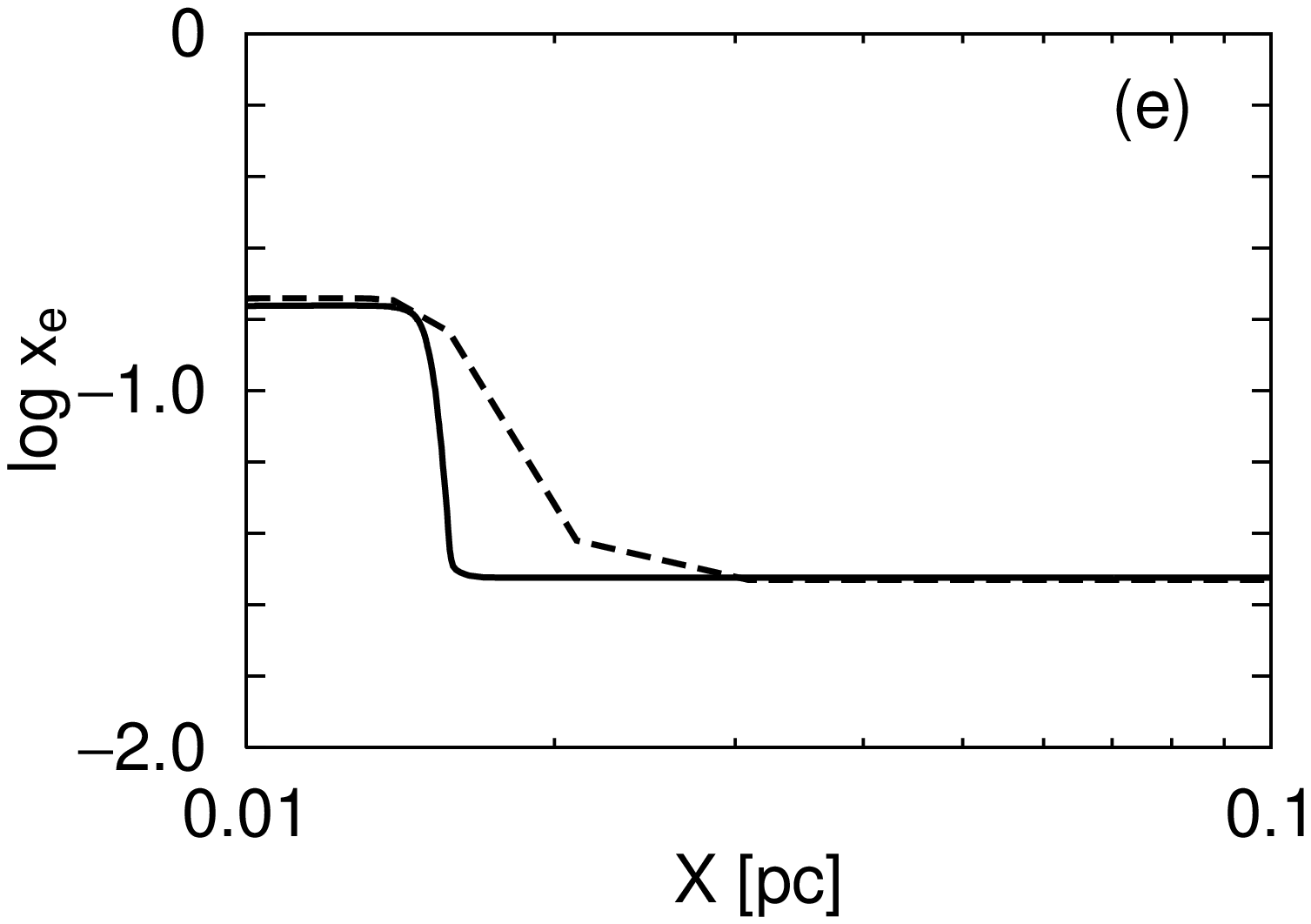}
\caption{Structure across the shock front.
The model W6 at $t=1\times 10^4$ yr is shown.
The figures show (a) pressure, (b) temperature, (c) number density of
hydrogen nuclei, (d) velocity, and (e) electron number fraction.
Dashed lines correspond to the same spatial resolution as figure 
\protect\ref{fig:WNM Layer}.
Solid lines denote the calculation with 320 times higher spatial
resolution than dashed lines. 
Dotted lines denote the calculation with 640 times higher resolution 
without cooling and heating.
}
\label{fig:shock front}
\end{figure}

In this subsection, we analyze the detailed structure across
shock front which propagates into WNM.
The same calculation as model W6 at $t=1\times 10^4$ are shown in
Figure \ref{fig:shock front}.
Dashed lines show the calculation 
which used the same initial grid spacing as model W6.
We also calculated the model with 320 times higher spatial resolution
(solid lines) to resolve detailed structure of the shock front.
Dotted lines denote the same calculation without cooling and heating 
processes.
The width of the shock front is on the order of collision mean free path,
$\ell=1/n\sigma_{col}\approx 0.0003/n$ pc.
For a strong shock without cooling, post-shock temperature
is determined by Rankine-Hugoniot relation as 
$T_2=(\gamma-1)mV^2_{\rm d}/2k_{\rm B}$.
On the other hands, 
shock temperature with cooling is estimated by the following equation: 
\begin{equation}
\frac{k_{\rm B}T_{\rm s}}{\gamma-1}+E_{\rm loss}=\frac{1}{2}mV^2_{\rm d},
\end{equation}
where 
$E_{\rm loss}=\int n(\Lambda-\Gamma)\, dt
\approx n\Lambda_{\rm Ly\alpha}(T_{\rm s}) \ell/V_{\rm d}$.
This analytic estimation provides $\log T_{\rm s} \ {\rm [K]}=4.8$ in
this model. 
Even the lowest resolution model can provide the same values of 
physical quantities in the post-shock region.  

Radiative precursor which is not treated in our models can change 
pre-shock temperature and ionization.
Shull \& McKee (1979) have calculated the ionization of the pre-shock
gas which is caused by the diffuse radiation from the post-shock
gas in their models of interstellar shock waves with $v_s=$ 40 -- 130 km/s. 
They have concluded that sufficiently fast shocks ($v_{\rm s}>110$ km/s)
completely ionize the pre-shock gas.
To study this effect,
we also calculated the model with fully ionized pre-shock gas (model I12),
and found no significant difference between the model I12 and W12.
Therefore we neglect the effect of radiative precursor in the rest of
this paper.

\subsection{Thermal Instability of Isobarically Cooling WNM}

Figure \ref{fig:n-Pw,n-Tw} shows the evolution of the fluid element 
which has the maximum density at each time step.
Density, temperature and pressure evolve on the dashed lines 
from left to right.
Solid and thick lines correspond to the thermally stable equilibrium. 
Dotted lines correspond to the thermally unstable equilibrium. 
The pre-shock gas is a
thermally stable WNM ($n=0.1 \, {\rm cm^{-3}}, \ T=8000$ K). 
In Appendix \ref{LINEAR}, 
we investigate the stability of isobarically contracting gas
by linear perturbation theory to analyze the evolution on the dashed lines.
We have used the post-shock values of hydrodynamic calculation as the
chemical compositions in the linear analysis
($x_{\rm e}=0.1$, $x_2=10^{-6.3}$, $x_{\rm CO}=10^{-23}$).
Because the chemical reaction timescale is longer than the cooling timescale.
Shaded area in Figure \ref{fig:n-Pw,n-Tw}b denotes 
thermally unstable region determined by the linear analysis.
Larger electron fraction makes cooling rate larger,
so that the unstable region is wider than the unstable region of
equilibrium gas.
This shaded region shows that thermal instability is inevitable
when WNM cools into CNM.

\begin{figure}[htbp]
\psbox[xsize=0.49\hsize]{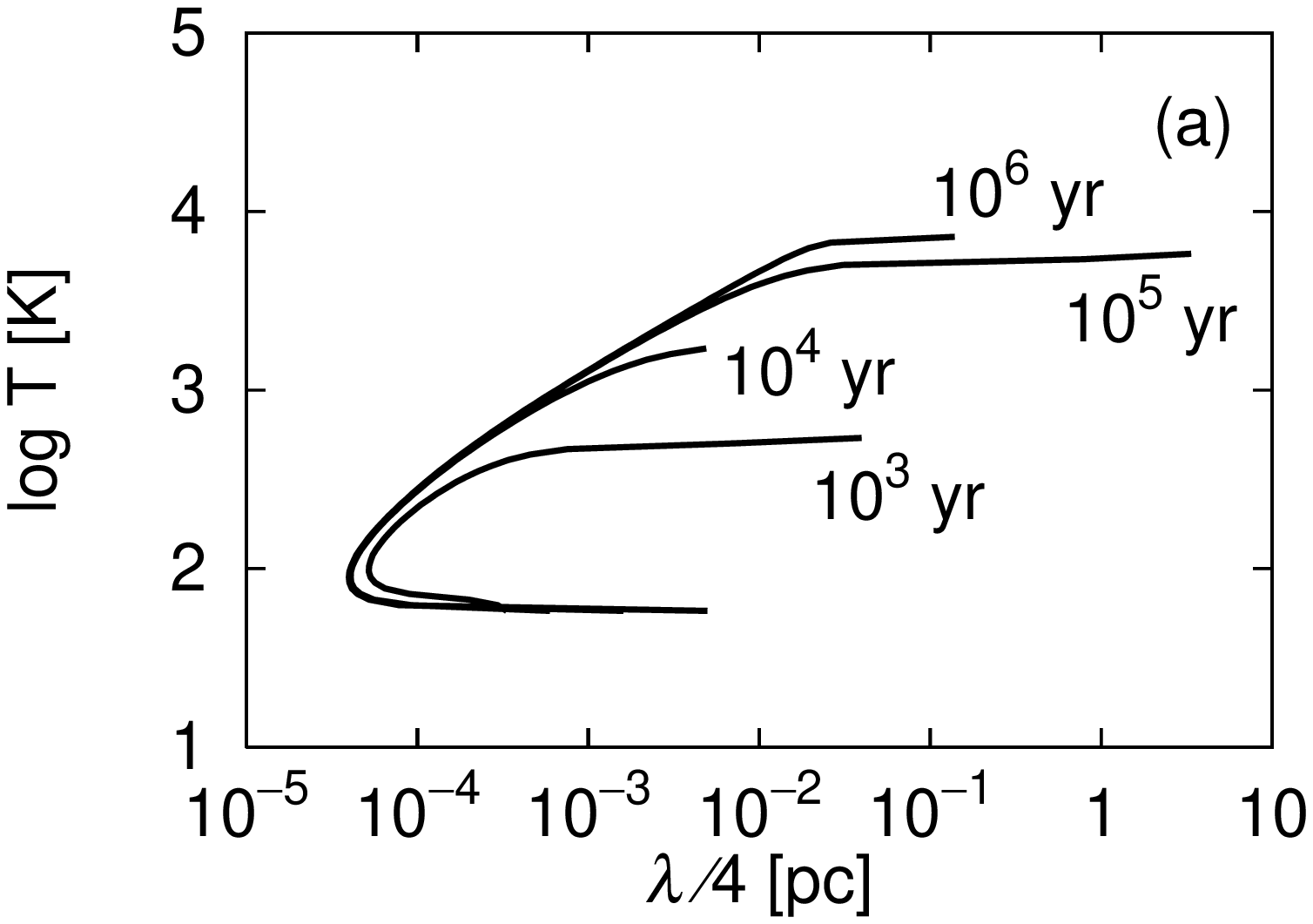}
\psbox[xsize=0.49\hsize]{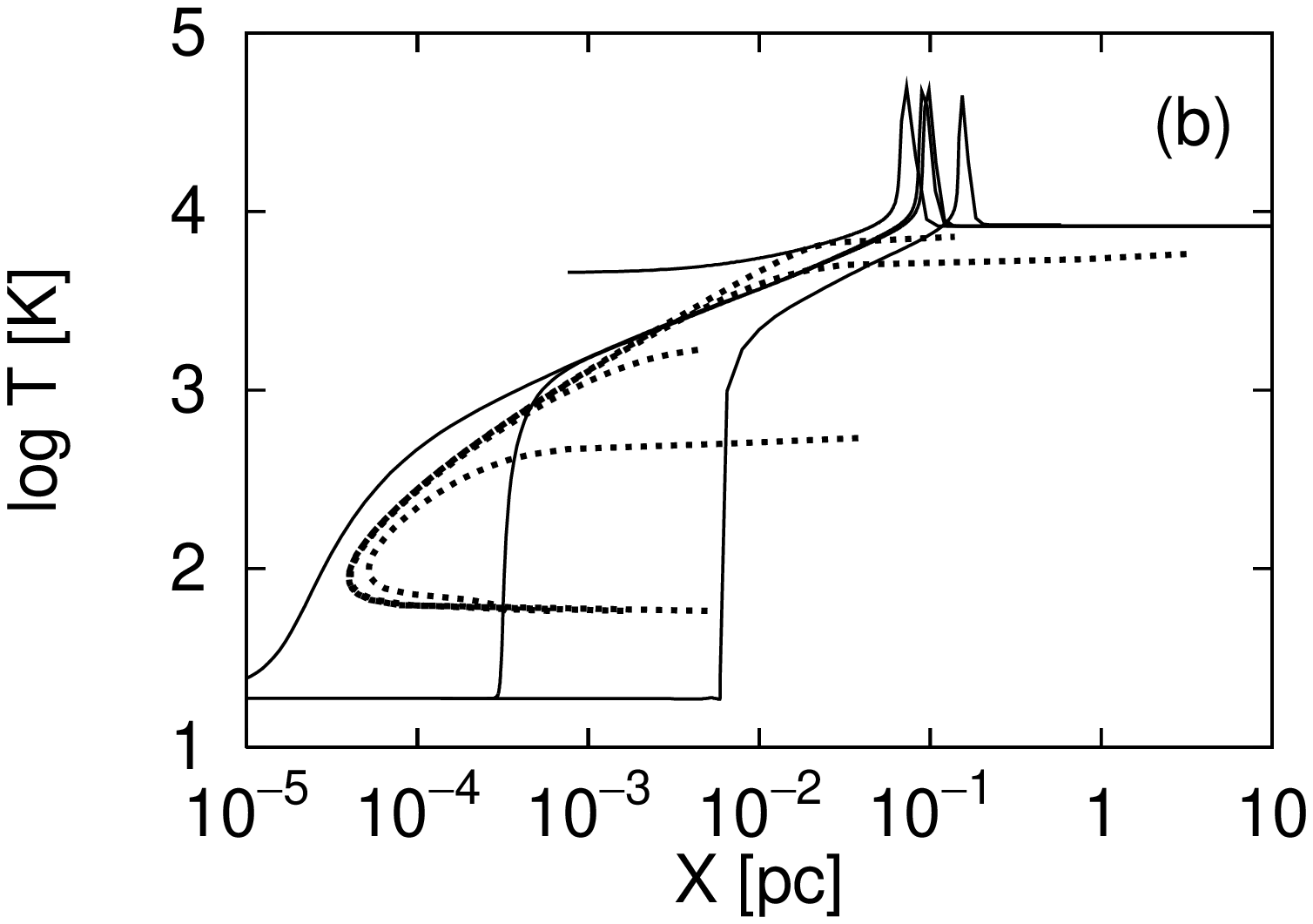}
\caption{The growth time of the instability.
The vertical axis denotes unperturbed temperature.
The value of the constant pressure, 
$P_{\rm c}/k_{\rm B}=5 \times 10^4 \,{\rm K/cm^3}$,
is adopted from the result of our non-linear calculation.
The density can be deduced from the relation, $n=P_{\rm c}/k_{\rm B}T$.
The horizontal axis denotes a quarter of perturbation wavelength.
Contours of growth time are depicted in this wavelength-temperature plane.
The shortest wavelength of unstable perturbation is
$\lambda_{\rm min}\sim 8 \times 10^{-5}\, {\rm pc} \approx 16 {\rm AU}$.
(b) We superpose the temperature evolution of our non-linear calculation 
(Figure \protect\ref{fig:WNM Layer}b)       
upon the growth rate contours (Figure \protect\ref{fig:growth43}a). }
\label{fig:growth43}
\end{figure}

Next, we discuss about the characteristic length and timescale of 
the thermal instability.
The linear growth rate of the thermal instability 
as a function of perturbation wavelength is calculated with 
a pair of given temperature and density.
The growth time of the instability is shown in Figure \ref{fig:growth43}.
In this figure, the vertical axis denotes the unperturbed temperature.
The value of the constant pressure, 
$P_{\rm c}/k_{\rm B}=5 \times 10^4 \,{\rm K/cm^3}$,
is adopted from the result of our non-linear calculation.
The density can be deduced from the relation, $n=P_{\rm c}/k_{\rm B}T$.
The horizontal axis denotes a quarter of perturbation wavelength.
Contours of growth time are depicted in this wavelength-temperature plane.
In Figure \ref{fig:growth43}b,
the temperature evolution of our non-linear calculation 
(Figure \ref{fig:WNM Layer}b) is superposed
upon the growth rate contours (Figure \ref{fig:growth43}a). 
Thermal instability occurs at temperature less than 7300 K.
The instability becomes drastic at temperature below 1000 K 
which corresponds to the growth time less than $10^4$ yr.
In this way we can understand the collapse of the layer in terms of
thermal instability.
The shortest wavelength of unstable perturbation 
is $\lambda_{\rm min} \sim 8\times 10^{-5} {\rm pc}\approx$ 16 AU.  
The thickness of the thermally-collapsed layer increases with time 
through the accretion of gas.

\subsection{Shock Propagation into CNM}\label{subsection:CNM}

\begin{figure}[htbp]
\psbox[xsize=0.49\hsize]{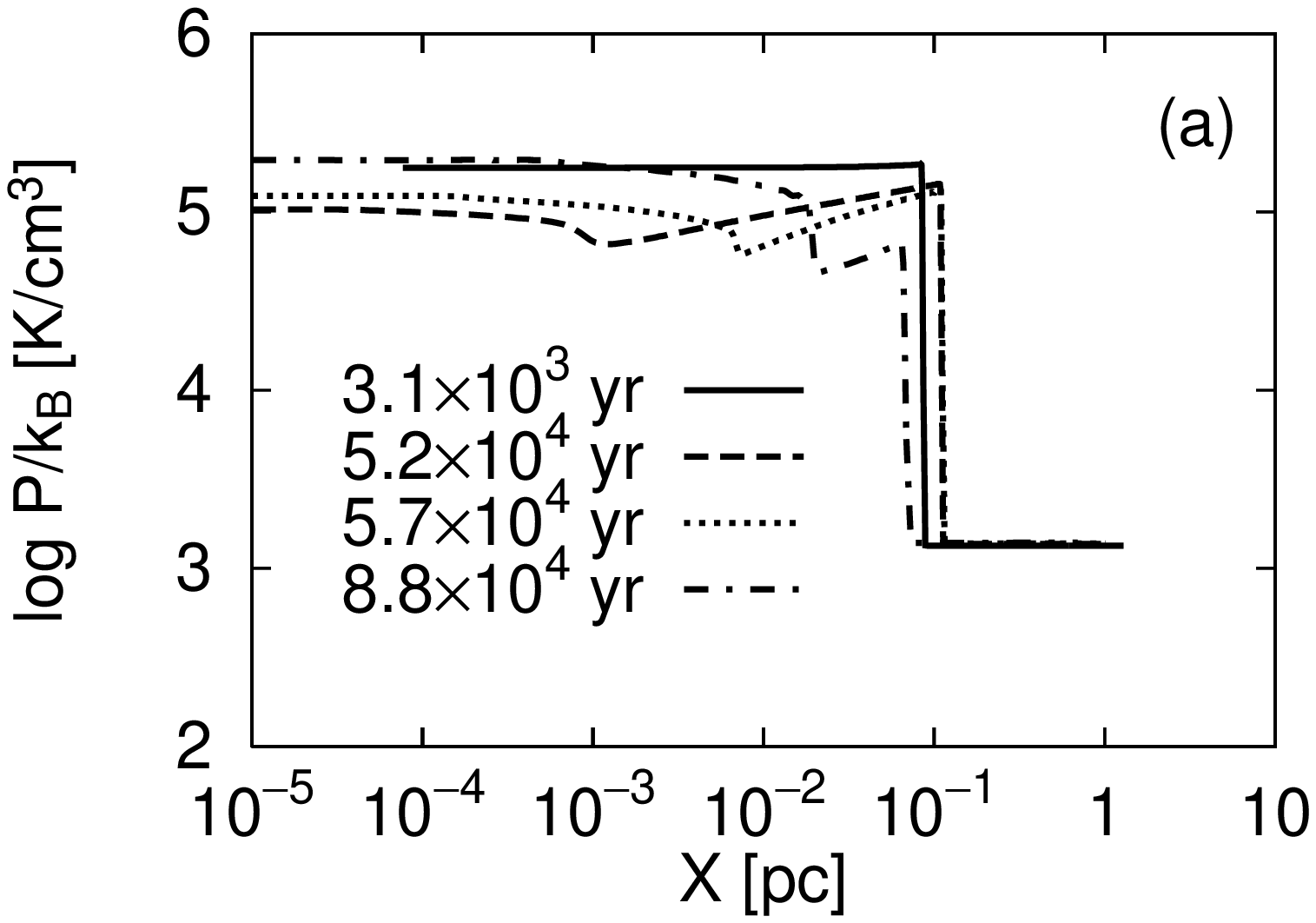}
\psbox[xsize=0.49\hsize]{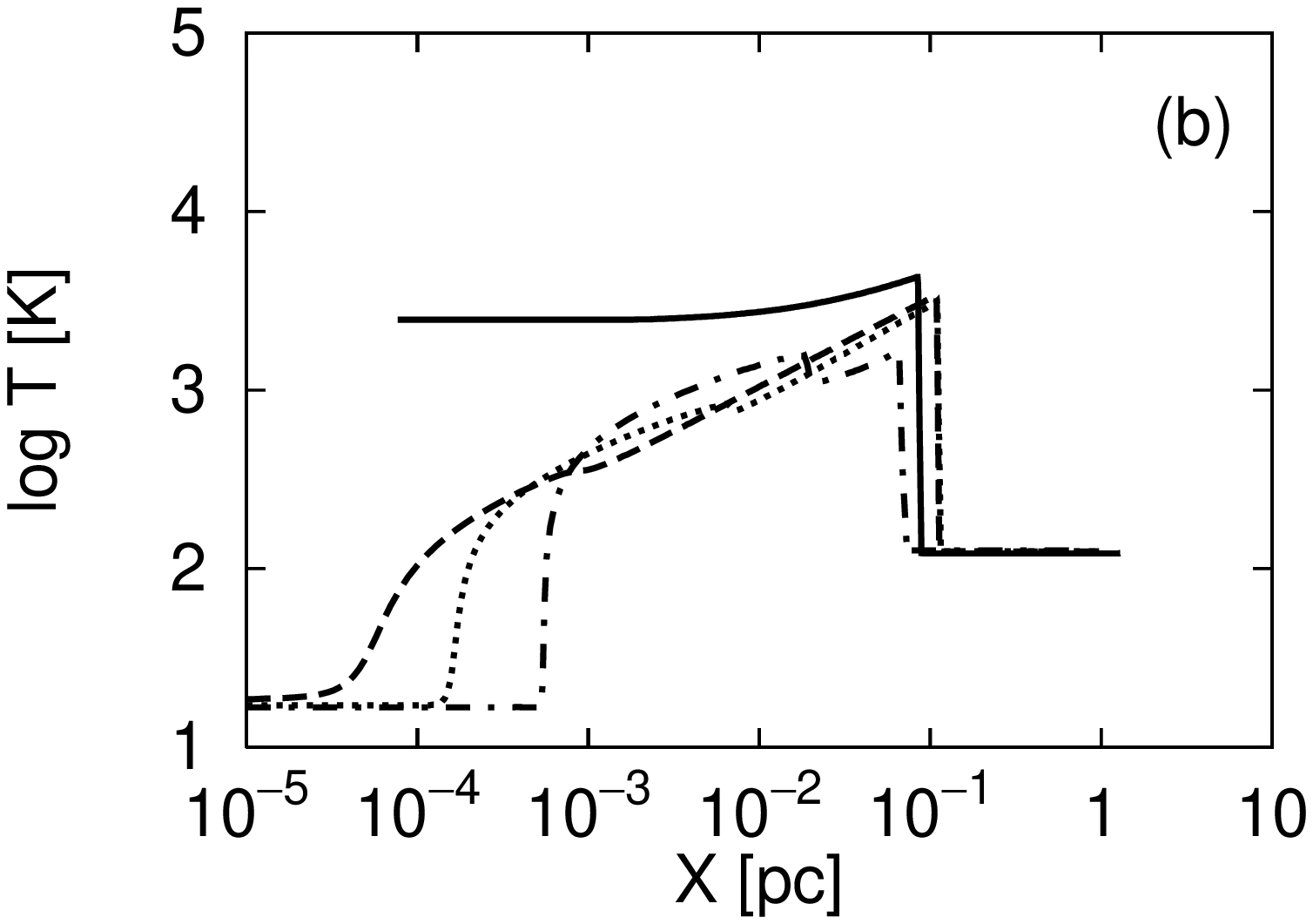}

\psbox[xsize=0.49\hsize]{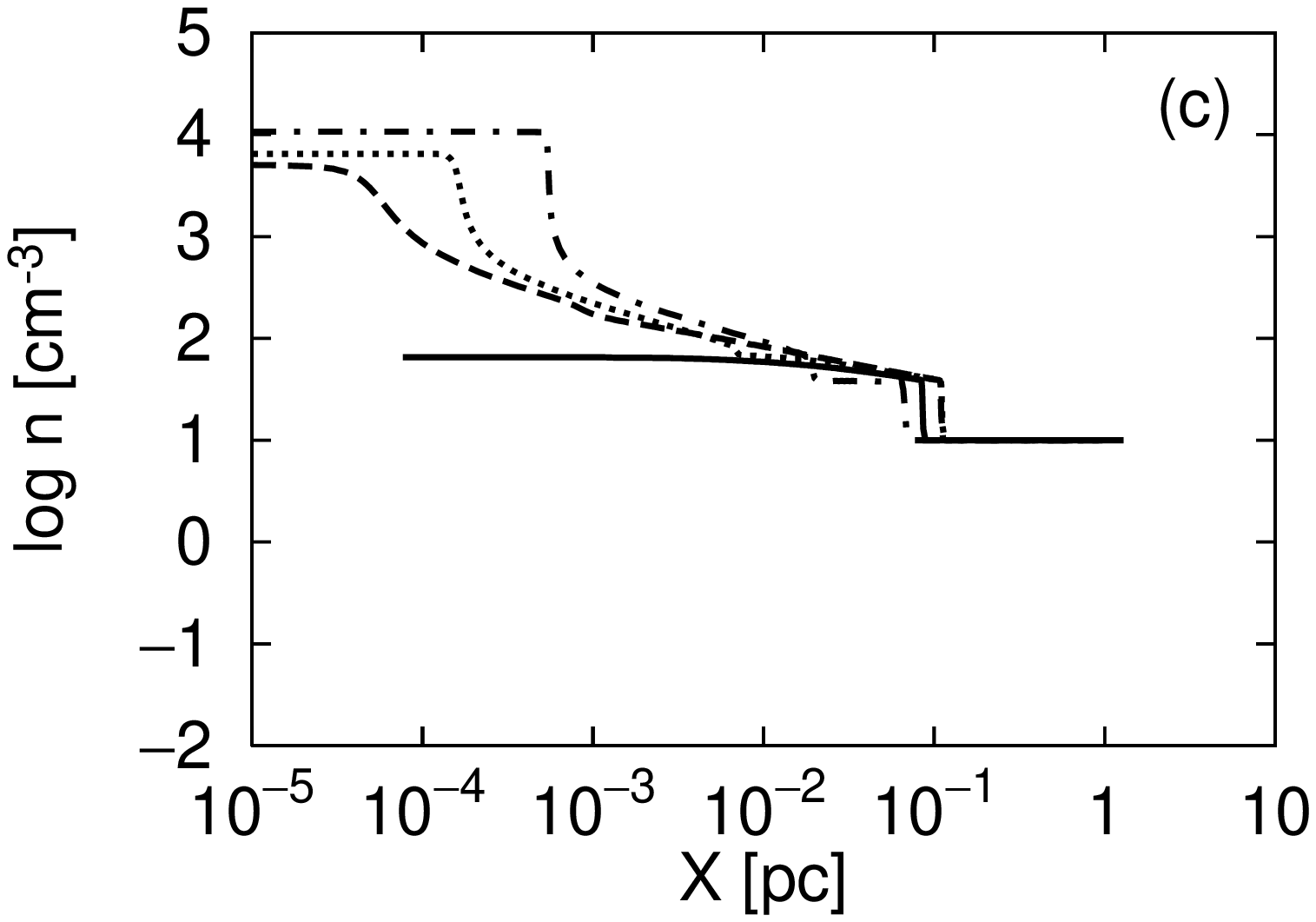}
\psbox[xsize=0.49\hsize]{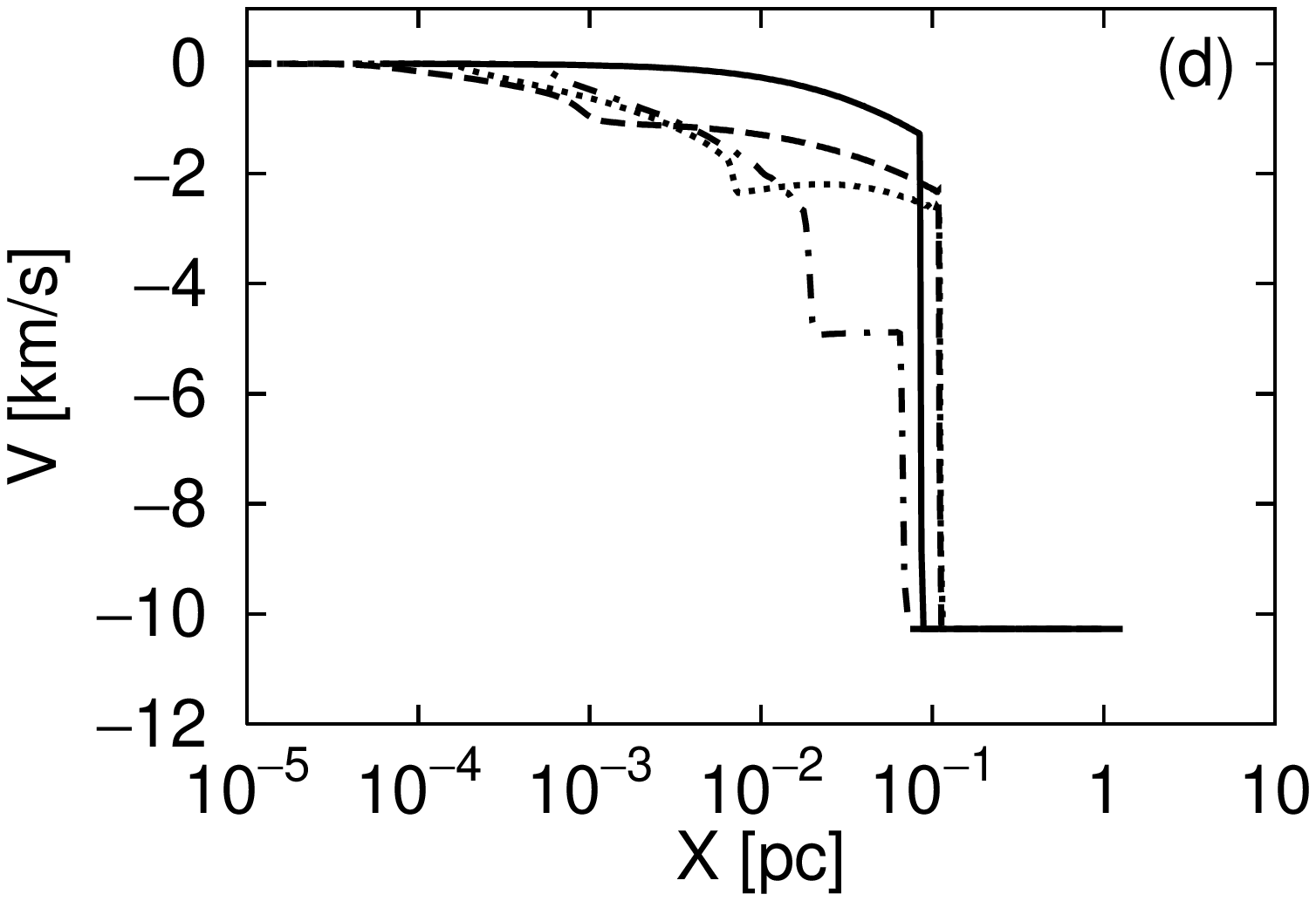}

\psbox[xsize=0.49\hsize]{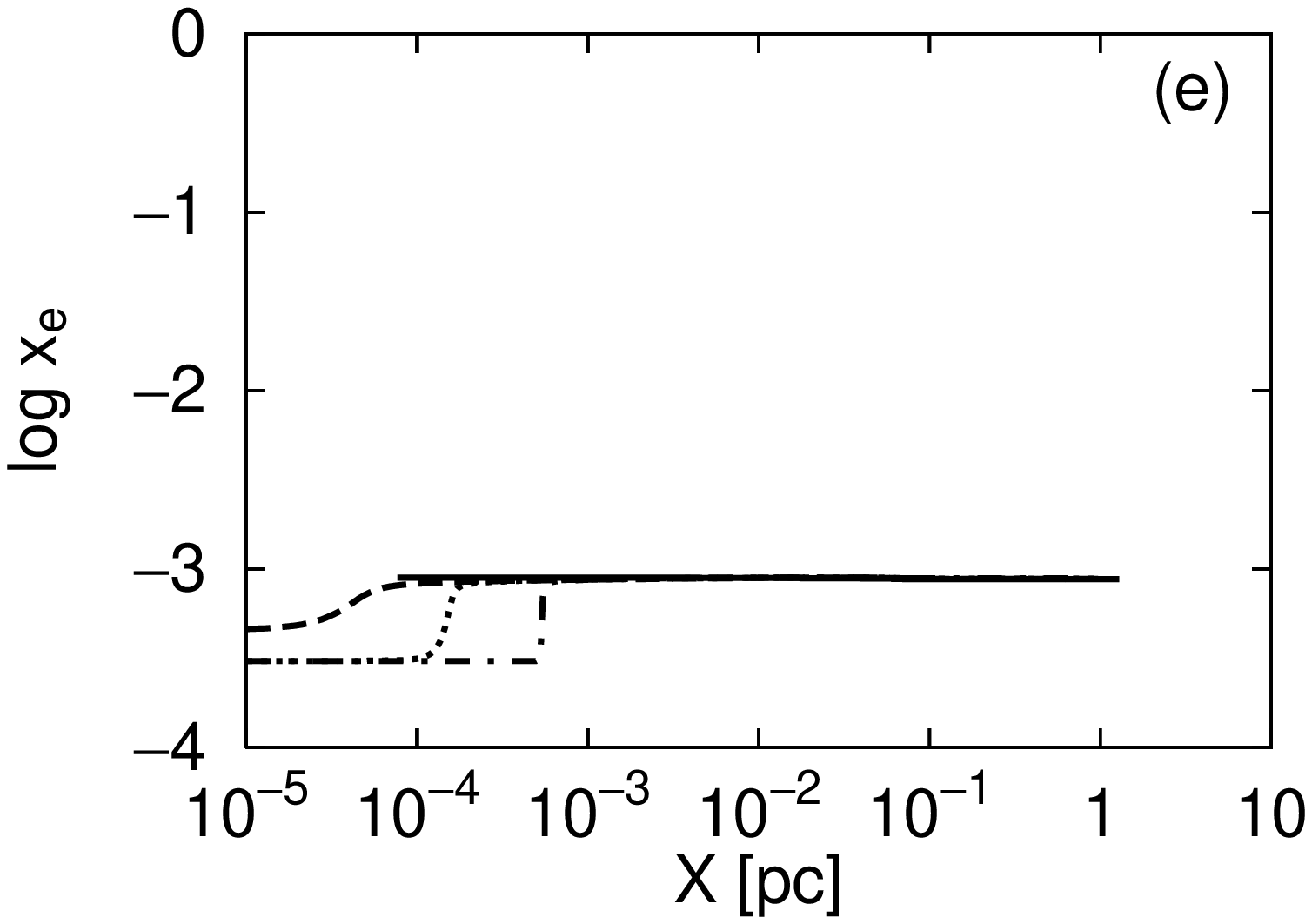}
\psbox[xsize=0.49\hsize]{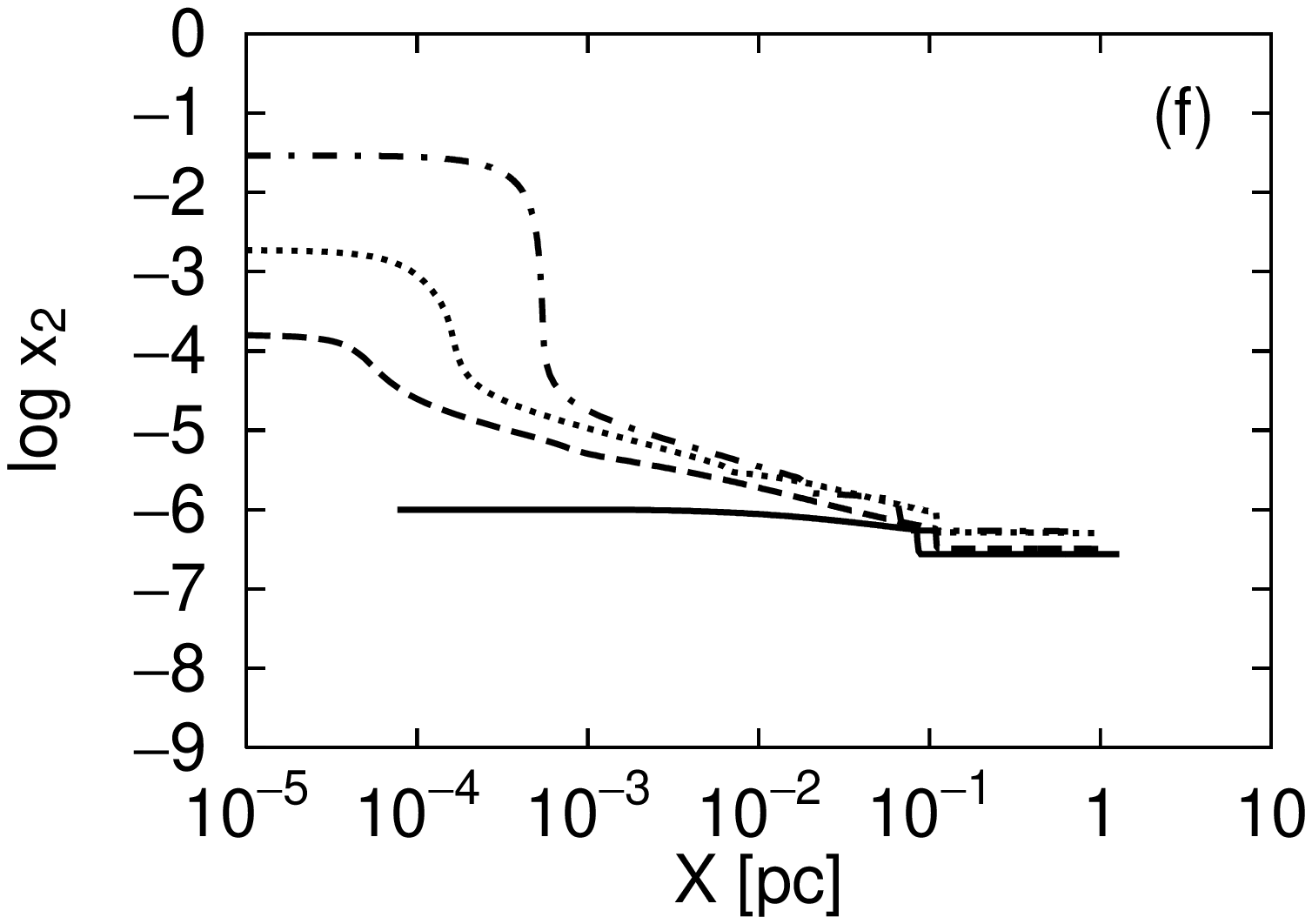}
\caption{The evolution of the shock propagation into the CNM.
The figures show (a) pressure, (b) temperature, 
(c) number density of hydrogen nuclei,
(d) velocity, (e) electron number fraction,
and (f) H$_2$ number fraction.
Solid lines denote the compressed layer at $t=3.1\times 10^3$ yr,
dashed lines at $t=5.2\times 10^4$ yr, 
dotted lines at $t=5.7\times 10^4$ yr, 
and dot-dash lines at $t=8.8\times 10^4$ yr.}
\label{fig:CNM Layer}
\end{figure}

In this section, we analyze the model C10 where shock wave propagates
into CNM.
Although Smith (1980) has done similar calculation,
his calculation does not have sufficient dynamic range to resolve 
thermally collapsing layer which is of our interest.
Figure \ref{fig:CNM Layer} shows the results of our calculation
where Mach number $M=10$ 
which corresponds to the velocity difference of 10 km/s.
Snapshots at $t=3.1\times 10^3$, $5.2\times 10^4$, $5.7\times 10^4$,
and $8.8\times 10^4$ yr are shown. 
The shock front is almost adiabatic.
The post-shock pressure can be estimated by Rankine-Hugoniot relation as
follows:
\begin{equation}
\frac{P_2}{P_1}=1+\frac{\gamma(\gamma+1)}{4}M^2+\frac{\gamma}{4}M
\sqrt{(\gamma+1)^2M^2+16},
\end{equation}
where M is the Mach number in the center of mass frame.
When M equals to 10, the post-shock pressure is 224 times higher than the
ambient pressure.
Details of post-shock structure is more complicated than 
that in the WNM case.
The post-shock gas, of which temperature and density are determined by
adiabatic shock condition, is thermally unstable 
(see Figure \ref{fig:n-Pc,n-Tc}b).
The shock front decelerates after post-shock layer collapsed.
After thermally collapsed layer is formed, 
accretion shock is formed inside the shock-compressed layer.
Accretion shock propagates into compressed layer.
In the highest density region at final time step,
4.9\% of the hydrogen is in H$_2$ and 
0.02\% of the carbon is in CO.
Note that the abundances of these molecules are still increasing
because the 
H$_2$
formation timescale is much longer than collapse timescale
$t\simeq 10^5$ yr (see Figure \ref{fig:equilibrium}d).

\subsection{Thermal Instability of Isobarically Cooling CNM}

\begin{figure}[htbp]
\psbox[xsize=0.49\hsize]{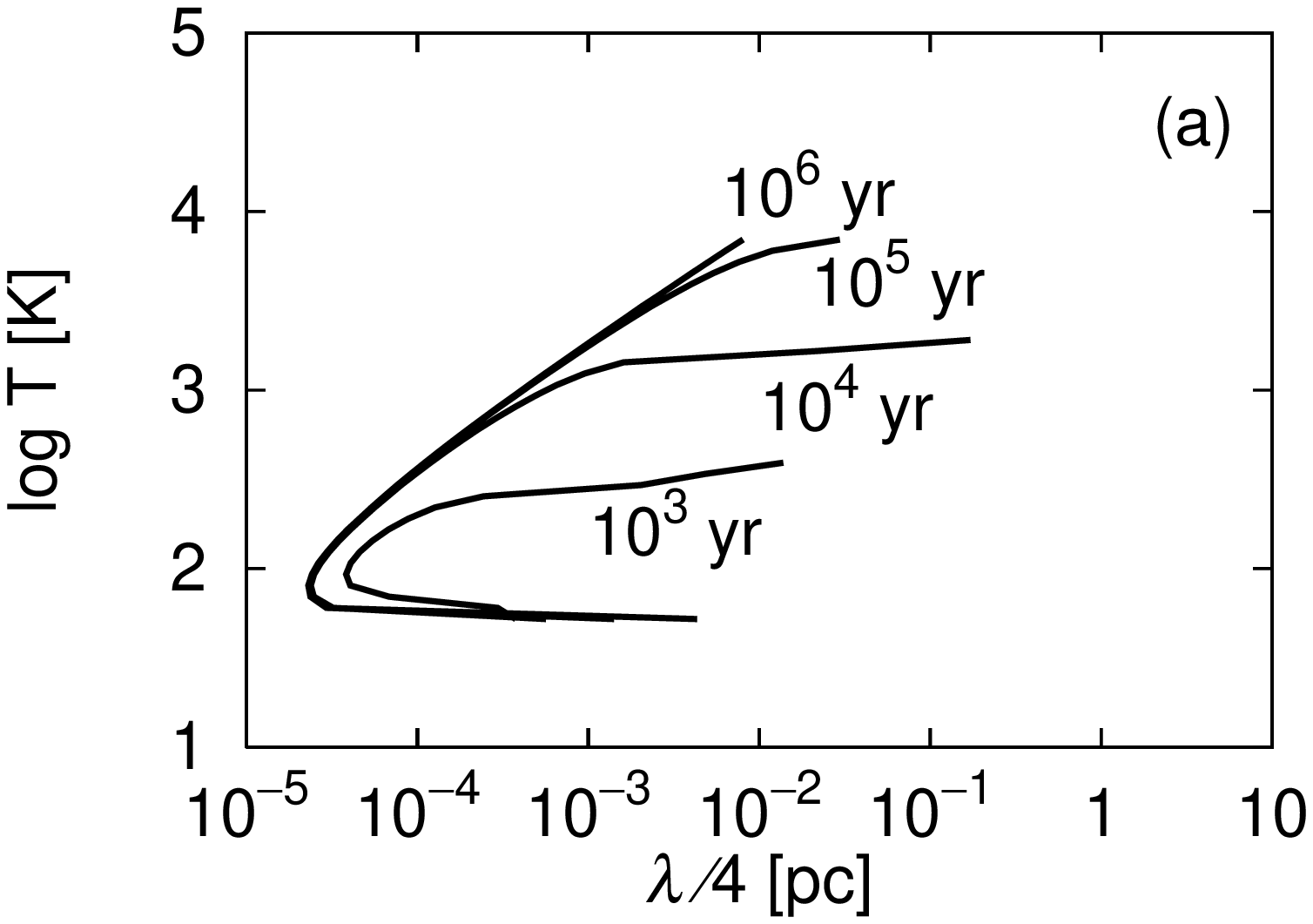}
\psbox[xsize=0.49\hsize]{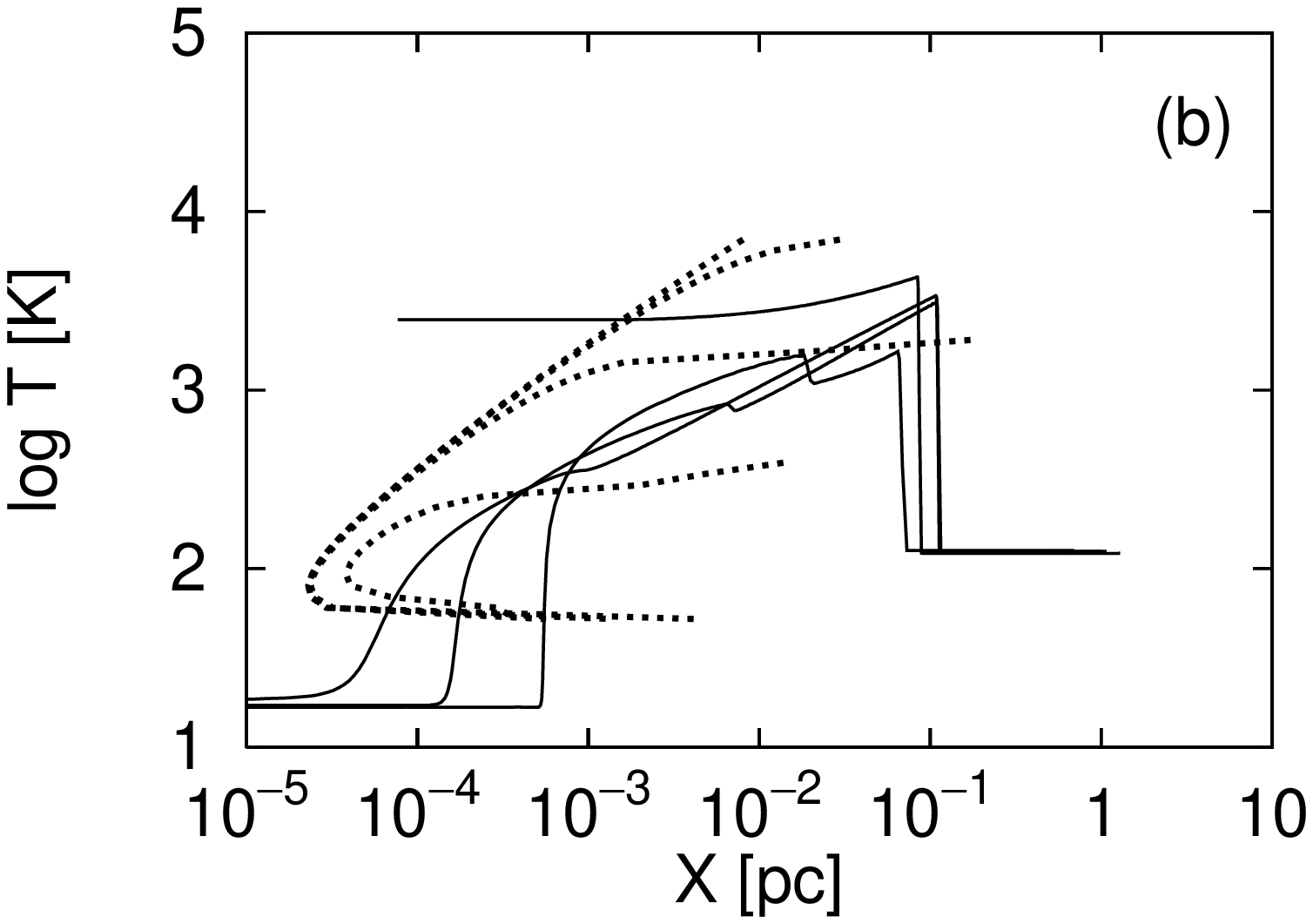}
\caption{The growth time of the instability.
The vertical axis denotes unperturbed temperature.
The value of the constant pressure, 
$P_{\rm c}/k_{\rm B}=2 \times 10^5 \,{\rm K/cm^3}$,
is adopted from the result of our non-linear calculation.
The density can be deduced from the relation, $n=P_{\rm c}/k_{\rm B}T$.
The horizontal axis denotes a quarter of perturbation wavelength.
Contours of growth time are depicted in this wavelength-temperature plane.
The shortest wavelength of unstable perturbation is 
$\lambda_{\rm min}\sim 2.5 \times 10^{-5}\, {\rm pc}\approx 5 {\rm AU}$.
(b) We superpose the temperature evolution of our non-linear calculation 
(Figure \protect\ref{fig:CNM Layer}b)       
upon the growth rate contours (Figure \protect\ref{fig:growth46}a). }
\label{fig:growth46}
\end{figure}

Figure \ref{fig:n-Pc,n-Tc} shows the evolution of the fluid element
which has the maximum density at each time step.
Shaded area of Figure \ref{fig:n-Pc,n-Tc}b denotes thermally unstable region.
In this case of linear analysis (Appendix \ref{LINEAR}), 
chemical compositions in thermal and chemical
equilibrium states of each densities are used. 
This Figure shows how thermally stable CNM becomes unstable.
The growth time of the instability is shown in Figure \ref{fig:growth46}.
In Figure \ref{fig:growth46}b,
we superpose the temperature evolution of our non-linear calculation 
(Figure \ref{fig:CNM Layer}b)       
upon the growth rate contours (Figure \ref{fig:growth46}a). 
The shortest wavelength of unstable perturbation is 
$\lambda_{\rm min} \sim 2.5\times 10^{-5} {\rm pc}\approx$ 5 AU.  

\subsection{Results from Various Initial Parameters}

\begin{figure}[htbp]
\psbox[xsize=.49\hsize]{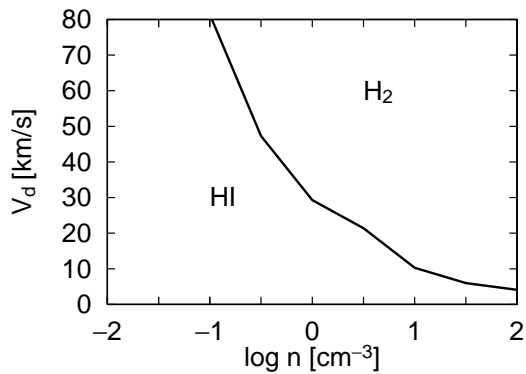}
\caption{
The chemical abundance of the thermally-collapsed layers
in the one-dimensional hydrodynamic calculation.
The horizontal axis denotes the initial number density.
The vertical axis denotes the velocity difference.
Solid lines denote the 8 \% boundary of H$_2$ number fraction.}  
\label{fig:chemical abundance}
\end{figure}

Calculations are done 
at nine different initial densities ($10^{-2}<n<10^2$),
and eleven velocity differences ($2<$ Mach number $<12$). 
The column density is $10^{20} \ {\rm cm^{-2}}$. 
We follow the calculation until $t=10^5$ -- $10^6$ yr.
The number fraction of H$_2$ in the highest density region
is shown in Figure \ref{fig:chemical abundance}. 
The horizontal axis denotes the initial number density of hydrogen nuclei.
The vertical axis denotes the velocity difference.
Solid lines denote the 8 \% boundary of H$_2$ number fraction.  
Higher velocity produces more abundance of H$_2$ 
in the thermally-collapsed layer.  
Note that the fragmentation of the thermally-collapsed layer will change
the column density of the fragments, which change the efficiency of 
H$_2$ self-shielding.
Therefore, the amount of H$_2$ in the fragments
would be different from the value obtained in this paper.
To determine the realistic abundance of H$_2$,
two or three-dimensional hydrodynamic calculation is required.

\section{DISCUSSION}

Our hydrodynamic calculation shows that thermal instability makes the
collapsed layer with considerable amount of H$_2$.
In this section, we discuss the consequence of this instability.

\subsection{Fragmentation of Thermally-Collapsed Layer}

The shock propagation into ISM can make the thermally-collapsing gas
inside the shock-compressed layer.
The initial thickness of the thermally-collapsed layer is 
dozens of AU and agrees with the prediction by the linear analysis. 
The thickness of the thermally-collapsed layer increases with time
through the accretion of gas.
The thermally collapsing layer is dynamically unstable
also in the Y- and Z-direction, so that perturbations 
in the layer will grow and lead to fragmentation of the layer.
We expect that this layer will break up into very small cloudlets
which have different translational velocities.
This velocity dispersion of cloudlets produced by the passage of SN shocks
may make an origin for the observed interstellar ``turbulent'' 
velocities.
Extremely high pressure in the tiny-scale structure observed in
\ion{H}{1} clouds (Heiles 1997) is consistent with our shock propagation
model.

\subsection{Molecular Clouds in The Galaxy}

It seems difficult to detect these cloudlets in emission lines 
of CO molecules (see e.g., Liszt \& Wilson 1993, Liszt 1994). 
Many shell-like or filamentary structures in the \ion{H}{1} 21-cm 
observation maps 
(Hartmann \& Burton 1997)
are reminiscent of overlapping shells of old SNRs. 
If these structures in \ion{H}{1} maps are really the 
results of SNR shocks,  
many tiny molecular cloudlets should be hidden in the shell 
and should be ubiquitous in the Galaxy. 
McKee \& Ostriker (1977) showed that the recycling SNRs in the 
galaxy would overlap before they are dissipated.
The overlapping filamentary region may produce larger clouds.

\section{SUMMARY}

We have done one-dimensional hydrodynamic calculations
for the propagation of a strong shock wave into WNM and CNM
including detailed thermal and chemical processes. 
Our results show that the shock propagation into WNM can make thin and
dense H$_2$ layer by the thermal instability inside the
shock-compressed layer.
The shock propagation into CNM can also make the thermally-collapsed layer.
We predict that this thermally-collapsed layer will fragment
into small molecular cloudlets.
Our subsequent work which includes two or three-dimensional
calculation is aimed to explore 
how these cloudlets have velocity dispersion, and form larger clouds.

\acknowledgments
We are grateful to the anonymous referee for helpful comments, 
which have improved the manuscript.
We would like to thank Shoken M. Miyama for discussions and 
continuous encouragement.

\appendix
\section{THERMAL PROCESSES}\label{THERMAL}
In this Appendix, thermal and chemical processes in our
calculation are summarized.
We included the formation and cooling of H$_2$ and CO in addition to the 
processes described by W95.

\subsection{Heating and Cooling Processes}
\subsubsection{Photoelectric Heating from Small Grains and PAHs}
The photoelectric emission from small grains and polycyclic aromatic 
hydrocarbons (PAHs) induced by FUV photons is an important mechanism
for heating a diffuse interstellar gas cloud.
We use the results of Bakes \& Tielens (1994, hereafter BT)
to calculate the photoelectric heating.
In their model, grains are distributed with a 
Mathis, Rumpl, \& Nordsieck (1977, hereafter MRN)
power-law distribution in size, $n(a)da\propto a^{-3.5}da$, 
between 3 and 100 \AA \, in radius.
BT calculated the charge distribution for each particle size taking the
photoionization and the electron and positive ion recombination into account 
and found the total photoelectric heating from the distribution of particles.
The resulting heating rate is given by

\begin{equation}
\Gamma_{\rm pe}=1.0\times 10^{-24}\epsilon G_0\,
{\rm ergs \,s^{-1}},
\label{eq:grain heating}
\end{equation}
where $\epsilon$ is the heating efficiency
and $G_0$ is the incident FUV field normalized to 
the local interstellar value
($=1.6\times 10^{-3} {\rm ergs \, cm^{-2}s^{-1}}$)
estimated by Habing (1968).
BT provide a simple fit to $\epsilon$ as a function of
$G_0 T^{1/2}/n_{\rm e}$:

\begin{equation}
\epsilon =
\frac{4.9\times 10^{-2}}{1.0+[(G_0T^{1/2}/n_{\rm e})/1925]^{0.73}}
+\frac{3.7\times 10^{-2}(T/10^4)^{0.7}}
{1.0+[(G_0T^{1/2}/n_{\rm e})/5000]},
\end{equation}
where $n_{\rm e}$ is the electron density.

Grains and PAHs may also be an important gas coolant.
We treat the heating caused by the electron ejection 
(eq.[~\ref{eq:grain heating}]) and the recombination cooling separately, 
so that the net heating or cooling is the difference between the heating 
and cooling.
We use the fit to the recombination cooling provided by BT:
\begin{equation}
\Lambda_{\rm pe}=
4.65\times 10^{-30}T^{0.94}(G_0T^{1/2}/n_{\rm e})^{\beta}n_{\rm e}\,
{\rm ergs \ s^{-1}},
\end{equation}
with $\beta=0.74/T^{0.068}$.

\subsubsection{Ionization Heating by Cosmic Rays and Soft X-Rays}
When the UV flux is strongly attenuated by hydrogen photo-absorption,
cosmic rays provide a main heating mechanism.
We adopt a primary cosmic-ray ionization rate of 
$\zeta_{\rm CR}=1.8\times 10^{-17} {\rm s^{-1}}$ 
(including the ionization of He) following W95.
Primary electrons ejected by cosmic-ray ionization have a larger kinetic 
energy than that of thermal electrons. 
This primary electron heats surrounding gas by secondary ionization.
The total heating rate is given by

\begin{equation}
\Gamma_{\rm CR}=\zeta_{\rm CR}E_h(E,x_{\rm e}),
\end{equation}
where the function $E_h(E,x_{\rm e})$ gives the
heat deposited for each primary electron of energy $E$
(Shull \& Van Steenberg 1985).

Soft X-rays also ionize and heat ISM.
W95 calculated the heating and ionization rates using 
the observed diffuse X-ray spectrum and photo-ionization cross section of
H, He, and other trace elements.
The analytic fits are given in their paper.
Soft X-ray is attenuated by photo-absorption of neutral hydrogen atom.
We use the value of absorbing column density 
$10^{19-20}\,{\rm cm^{-3}}$ as a standard \ion{H}{1} clouds.

\subsubsection{Formation and Photodissociation Heating from H$_2$}
When a hydrogen molecule is photodissociated the chemical potential of the gas
is raised by the dissociation energy, 4.48 eV.
In addition to this, a small amount of kinetic energy
is deposited to the photodissociated atoms.
According to Hollenbach \& McKee (1979, hereafter HM79),
the heating rate is

\begin{equation}
\Gamma_{\rm UV}=9 R_{\rm pump}
\left\{(2.2{\rm eV})\left[1+n_{\rm cr}(H_2)/n\right]^{-1}\right\},
\end{equation}

where 

\begin{equation}
n_{\rm cr}=\frac{10^6T^{-1/2}}{1.6x_{\rm H}\exp[-(400/T)^2]
+1.4x_2\exp \{-[12000/(T+1200)]\}} \ \ {\rm cm^{-3}},
\end{equation}
and the rate coefficient $R_{\rm pump}$ is given in 
Appendix \ref{Rpump}.

When a hydrogen molecule forms through associative detachment
or catalytic reaction,
it is in a highly excited rovibrational state.
Only a small fraction of the formation energy is
available for heating the gas, while most of the energy escapes the cloud
via line emission.
We use the heating rates from HM79
as follows:
\begin{equation}
\Gamma_{\rm gr}=Rx_{\rm H}
\left\{0.2+4.2[1+n_{\rm cr}({\rm H_2})/n]^{-1}\right\},
\end{equation}
and the rate coefficient $R$ is given in Appendix \ref{Rpump}.

\subsubsection{Atomic Line Cooling}
At the high temperatures ($T\gtrsim 8000$ K) collisional excitation 
of hydrogen Ly$\alpha$ can contribute to the cooling.
We also include low-lying metastable transitions of 
\ion{C}{2} (2326\AA), \ion{O}{1} (6300\AA), \ion{Fe}{2} (5.3 \micron), 
and \ion{Si}{2} (2240\AA),
using rates from Hollenbach \& McKee (1989, hereafter HM89).

Collisional excitation of the fine-structure lines of
\ion{C}{2} (158 \micron) and \ion{O}{1} (63 \micron) is a dominant gas coolant 
at temperatures $T\lesssim 8000$ K.
Additional cooling is provided by the 
fine-structure transitions of the ground electronic state
of \ion{Fe}{2} (26 \micron), \ion{Si}{2} (35 \micron) (HM89).

We adopt trace element abundance following W95; 
$x_{\rm He}$ = 0.1;
$x_{\rm O}$ = 4.6$\times 10^{-4}$; 
$x_{\rm C}$ = 3$\times 10^{-4}$; 
$x_{\rm Si}$ = 3.55$\times 10^{-6}$; 
$x_{\rm Fe}$ = 7.08$\times 10^{-7}$. 
Several trace elements, C, Si, and Fe, have ionization potentials less than 
13.6 eV and thus are susceptible to ionization by the interstellar UV field.
For convenience, the level of ionization of these trace elements is
assumed to be in singly charged.
On the other hands, 
oxygen remains neutral by charge exchange reaction to hydrogen, 
because oxygen has almost the same ionization potential as hydrogen's.

\subsubsection{H$_2$ Cooling}
At temperatures above 500 K, rovibrational lines of H$_2$
contribute to the cooling function.
Following HM79 we adopt
\begin{equation}
\Lambda_{\rm H_2}\approx n({\rm H_2})\left[
 x_{\rm H}\frac{L_{\rm vr}^{\rm H}({\rm LTE})}{1+n^{\rm H}_{\rm cr}/n}
+x_{\rm 2}\frac{L_{\rm vr}^{\rm H_2}({\rm LTE})}{1+n^{\rm H_2}_{\rm cr}/n}
\right],
\end{equation}
where $n_{\rm cr}$ is the critical density defined as follows:

\begin{equation}
n_{\rm cr}^{\rm H,H_2}\equiv 
\frac{L_{\rm vr}^{\rm H,H_2}({\rm LTE})}{L_{\rm vr}^{\rm H,H_2}(n\to 0)}n,
\end{equation}
which depends only on temperature, not on density.
We use the function $L_{\rm vr}^{\rm H,H_2}({\rm LTE})$,
$L_{\rm vr}^{\rm H,H_2}(n \to 0)$ from HM79.
We update the function $L^{\rm H}_{\rm r}(n \to 0)$ from Galli \& Palla (1998).
We assume an ortho -- H$_2$ to para -- H$_2$ ratio of 3:1.

\subsubsection{CO Cooling}
Rotational line emission from CO is a major
low-temperature cooling agent for molecular clouds.
The CO rotational cooling is derived from McKee et al. (1982).
The optically thin cooling function is

\begin{equation}
\Lambda_{\rm CO}({\rm rot})
=\frac{4(k_{\rm B}T)^2A_0}{E_0[1+(n_{\rm cr}/n)+1.5(n_{\rm cr}/n)^{1/2}]},
\end{equation}
where $A_0=9.7\times 10^{-8}{\rm s^{-1}}$, $E_0/k_{\rm B}=2.76$ K , 
$n_{\rm cr}=3.3\times 10^6 T_3^{3/4}\, {\rm cm^{-3}}$
for ${}^{12}$CO (McKee et al. 1982).

The CO vibrational cooling at the lowest level is
\begin{equation}
\Lambda_{\rm CO}({\rm vib})=
\sum_{\rm H,H_2}\gamma_{01}^{\rm H,H_2}\Delta E_{10},
\end{equation}
where $\Delta E_{10}=3080 {\rm K}/k_{\rm B}$.
The vibrational rate coefficients for transitions to
the v=1 state of CO were taken from HM89.

\subsubsection{Atomic and Molecular Collisions with Dust Grains}
The cooling of the gas by cooler dust grains is important at
high densities $n\gtrsim 10^5 \, {\rm cm^{-3}}$.
The cooling rate per hydrogen nuclei is

\begin{equation}
\Lambda_{\rm gr}=1.2\times 10^{-31}n
\left(\frac{T}{\rm 1000K}\right)^{1/2}
\left(\frac{100 \,{\rm \AA}}{a}\right)^{1/2}
\times[1-0.8\exp(-75/T)](T-T_{\rm gr}) \ {\rm ergs \ s^{-1}},
\end{equation}
where $T_{\rm gr}$ is an effective grain temperature, averaged over the
assumed MRN size distribution (HM89).
Since the gas generally collides with the smaller grains,
we take $T_{\rm gr}$ to be the temperature 8 K of grains of radius 
$a=100 \,{\rm \AA}$.

\subsection{Chemical Reactions}
For the range of temperatures and densities considered, the recombination time
scale is longer than the cooling timescale
(See Figure \ref{fig:equilibrium}d).
The residual electrons make the cooling rates larger than 
that of equilibrium abundance.
Therefore, it is imparative to treat chemical non-equilibrium processes.

The main coolant in dense clouds is supposed to be CO, 
so that CO formation should be taken into account. 
Although H$_2$ is minor coolant, H$_2$ enables the formation of CO.
Thus, we should take into account the formation of H$_2$. 

\subsubsection{Ionization and Recombination of Hydrogen Atom}
Photoionizing UV fields are attenuated by large absorption cross section of 
hydrogen atom.
The hydrogen atom is also ionized by cosmic rays and X-rays.
The total ionization rates per hydrogen atom are given 
in the Appendix of W95.
Collisional ionization is also important in shock-compressed layers,
and rate coefficient of this processes is 
$k_{\rm I}=5.8\times 10^{-9}T_4^{0.50}\exp(-15.8/T_4)\,
{\rm cm^{-3}s^{-1}}$(HM79).
The radiative recombination coefficient of hydrogen is given
by Shapiro \& Kang (1987).

\subsubsection{Formation and Dissociation of H$_2$}\label{Rpump}

H$_2$ is formed catalytically on dust grains
and by associative detachment; 
it is destroyed by collisional dissociation, 
by photodissociation via the Lyman-Werner electronic transitions,
and by cosmic rays. 

Following Tielens \& Hollenbach (1985, hereafter TH85), the rate at which
H$_2$ is produced catalytically on dust grains is
\begin{equation}
R=0.5\bar{v}n_{\rm H}n_d\sigma_dS(T)\approx
6\times10^{-17}(T/300)^{0.5}n_{\rm H}nS(T)\,{\rm cm^{-3}s^{-1}},
\end{equation}
\begin{equation}
S(T)=
[1+0.04(T+T_{\rm gr})^{0.5}+2\times 10^{-3}T+8\times 10^{-6}T^2]^{-1},
\end{equation}
where $\bar{v}$ is the mean thermal velocity of the atomic 
hydrogen, $n_d$ is the number density of dust grains, and
$\sigma_d$ is the average dust grain cross section.

H$_2$ can be formed by associative detachment of the 
H$^{-}$ ion,

\begin{equation}
{\rm H}+{\rm H}^{-}\to {\rm H_2}+e^{-}.
\end{equation}
We use the formation rates from HM79 to obtain an effective rate coefficient.

The shock waves can produce sufficiently high temperatures
so that collisional dissociation of H$_2$ should be considered.
The rate of dissociation of H$_2$ by H atoms is 
$k_{\rm D}=3.4\times 10^{-9}\exp(8000/T)\exp(-5.19\times 10^4/T)
\, {\rm cm^{-3}s^{-1}}$ (HM79).
The rate of dissociation of H$_2$ by electron impact with LTE
vibrational population is taken into account (Stibbe \& Tennyson 1999). 
Although these LTE reaction rates are over-estimates for H$_2$ dissociation,
there is no significant difference between models with and without 
collisional dissociation.
H$_2$ forms after the gas layer collapses,
because the timescale of H$_2$ formation is much larger than
cooling timescale. 
The temperature in collapsed layer is too low for H$_2$ to 
dissociate by collision. 
This is why collisional dissociation is ineffective
in our models.

We adopt the rate of photodissociation from TH85,
\begin{equation}
R_{\rm pump}=3.4\times 10^{-10}G_0\beta(\tau)\,{\rm s^{-1}},
\end{equation}
where $\beta(\tau)$ is the probability that
photons of the interstellar radiation field penetrate to optical depth $\tau$ 
in a plane parallel cloud layer.
Photodissociating UV fields are self-shielded by H$_2$,
and the optical depth $\tau$ is 

\begin{equation}
\tau=1.2\times 10^{-14}N({\rm H_2})\delta V_D^{-1},
\end{equation}
where $N({\rm H_2})$ is the column density of H$_2$ and
$\delta V_D$ is the Doppler line width in km s$^{-1}$,
and we adopt this value to be the sound velocity.
The analytic expression of $\beta(\tau)$ is given by TH85.

The rate at which cosmic rays destroy H$_2$
is estimated by HM89 to be 
2.29$\zeta_{\rm CR}$, where $\zeta_{\rm CR}$ is the 
primary cosmic-ray ionization rate for atomic hydrogen.

\subsubsection{Carbon Chemistry}
We include a simplified treatment of the conversion of 
singly ionized carbon C$^{+}$ to carbon monoxide CO
following Langer (1976) and Nelson \& Langer (1997).
The simplified chemical model assumes the direct conversion 
of C$^+$ to CO, or vice versa, without accounting explicitly 
for the intermediate reactions.
The equation describing the rate of production and destruction 
of CO is written
\begin{equation}
\frac{dn({\rm CO})}{dt}=
k_0n({\rm C^+})n\beta-\Gamma_{\rm CO}n({\rm CO})
\end{equation}
where $k_0=5\times 10^{-16}\,{\rm cm^3 s^{-1}}$, 
$\Gamma_{\rm CO}=10^{-10}G_0\ {\rm s^{-1}}$, 
and $\beta$ is defined by the expression

\begin{equation}
\beta=\frac{k_1x(\rm{O} \; \small{I})}
{k_1x(\rm{O} \; \small{I})+G_0[\Gamma_{\rm CHx}/n({\rm H_2})]}. 
\end{equation}
The definition of $\Gamma_{\rm CHx}$ is given by 
$\Gamma_{\rm CHx}=5\times 10^{-10}G_0$, and $k_1$ takes the 
value $k_1=5\times 10^{-10}$.
For CO photodissociation we do not treat the line self-shielding of CO
following Nelson \& Langer (1997),
because we are only interested in the beginning of CO formation.

\section{THERMAL INSTABILITY IN ISOBARICALLY CONTRACTING GAS}\label{LINEAR}
To study the characteristic length and timescale of the
thermally-collapsed layer,
we have done linear stability analysis.
Field (1965) analyzed the stability of uniform gas in thermal equilibrium.
Schwarz et al. (1972) analyzed the isochorically cooling gas without heating. 
Our non-linear calculation, however, shows that 
the shock-compressed layer is almost isobaric.
In this appendix, we analyze the stability of isobarically contracting gas.

\subsection{The Basic Equations}

We consider the gas with number density $n$, temperature $T$,
velocity $v$, and pressure $P$, 
where all variables are functions of position and time.
The governing hydrodynamic equations are

\begin{equation}
\frac{dn}{dt}+n\frac{\partial v}{\partial X}=0,
\end{equation}
\begin{equation}
m_{\rm H}n\frac{dv}{dt}=-\frac{\partial P}{\partial X},
\end{equation}
\begin{equation}
\frac{1}{\gamma-1}\frac{dP}{dt}
-\frac{\gamma}{\gamma-1}\frac{P}{n}\frac{dn}{dt}
=n\Gamma-n\Lambda+
\frac{\partial}{\partial X} \left(K\frac{\partial T}{\partial X}\right),
\end{equation}
\begin{equation}
P=nk_{\rm B}T.
\end{equation}

In this analysis,
we neglect He atoms for simplicity.
The effect of chemical evolution can be included in non-equilibrium 
cooling and heating function.
For the range of temperatures and densities considered, 
the chemical timescale is longer than the cooling timescale.
We use the non-equilibrium chemical fraction just after shock heating. 

\subsection{The Unperturbed State}

To express isobarically contracting background,
we use comoving coordinate $x$ related to real coordinate $X$,
\begin{equation}
x=X/a(t),
\end{equation}
where $a(t)$ is the contraction parameter.
In these contracting coordinates, the velocity field can be written as
\begin{equation}
v=\dot{a}x+v_1,
\end{equation}
where $v_1$ is the perturbed velocity relative to the contraction.

Lagrangian differentiation of physical variable $f$ is 
\begin{equation}
\frac{df}{dt}
=\frac{\partial f}{\partial t}+\left(v\frac{\partial}{\partial X}\right)f
=\left(\frac{\partial f}{\partial t}\right)_x
+\frac{1}{a}\left(v_1\frac{\partial}{\partial x}\right)f.
\end{equation}

In comoving coordinates, the one-dimensional hydrodynamic equations are
\begin{equation}
\dot{n}+\frac{\dot{a}}{a}n+\frac{1}{a}\frac{\partial}{\partial x}(nv_1)=0,
\end{equation}

\begin{equation}
\ddot{a}x+\dot{v_1}+\frac{\dot{a}}{a}v_1
+\frac{1}{a}\left(v_1\frac{\partial}{\partial x}\right)v_1
+\frac{1}{m_{\rm H}na}\frac{\partial P}{\partial x}=0,
\end{equation}

\begin{equation}
\dot{P}+\frac{1}{a}\left(v_1\frac{\partial}{\partial x}\right)P
-\gamma \frac{P}{n}
\left\{\dot{n}+\frac{1}{a}\left(v_1\frac{\partial}{\partial x}\right)n \right\}
+(\gamma-1)(n\Lambda-n\Gamma)
=\frac{\gamma-1}{a^2}\frac{\partial}{\partial x}
\left(K\frac{\partial T}{\partial x}\right).
\end{equation}
If we set the perturbed velocity $v_1$ equals to zero in the above equations,
the equations governing isobarically contracting background are
\begin{eqnarray}
\dot{n}+\frac{\dot{a}}{a}n=0, \label{eq:EOC-0} \\
\ddot{a}x+\frac{1}{m_{\rm H}na}\frac{\partial P_0}{\partial x}=0, \label{eq:EOM-0}\\
\gamma \frac{\dot{a}}{a}+(\gamma-1)\frac{\Lambda-\Gamma}{k_{\rm B}T}=0.
\label{eq:EOE-0}
\end{eqnarray}
Equation \ref{eq:EOC-0} leads $n\propto a^{-1}$.
Temperature is proportional to $a$
because unperturbed gas is isobaric. 
From equation \ref{eq:EOE-0}, we obtain the approximate solution for
$a(t)$ as
\begin{equation}
a(t)=a(0)(1-t/\gamma\tau_{\rm net}),
\end{equation}
where $\tau_{\rm net}$ is a net cooling time defined by 
$\tau_{\rm net}^{-1}=(\gamma-1)(\Lambda-\Gamma)/k_{\rm B}T$.
This approximation is valid as far as variations of density and
temperature remain small, i.e. $0<t\ll \gamma\tau_{\rm net}$.
Equation \ref{eq:EOM-0} gives $\partial P_0/\partial x=0$.
Thus, we consider uniform unperturbed state, $\partial n_0/\partial x=0$.
The isobarically cooling gas is characterized by a temperature
$T_0(t)$ and a number density $n_0(t)$ which are functions
of time, while the pressure $P_0$ is constant.

\subsection{The Linearized Equations}

We now examine the effect of small perturbations:
\begin{eqnarray}
n(x,t)&=&n_0(t)(1+n_1(t)\exp(i\kappa x)), \\
P(x,t)&=&P_0(1+P_1(t)\exp(i\kappa x)), \\
T(x,t)&=&T_0(t)(1+T_1(t)\exp(i\kappa x)),  \\
v_1(x,t)&=&v_1(t)\exp(i\kappa x).
\end{eqnarray}

\begin{figure}[htbp]
\psbox[xsize=0.49\hsize]{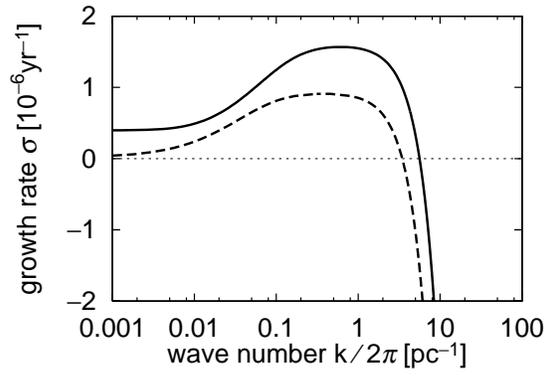}
\caption{Dispersion relation for the condensation mode of 
the isobarically contracting uniform gas
with $n=10^{0.25}$, $T=10^{3.0}$ (solid line).
The case of thermal equilibrium unperturbed state is also shown 
(dashed line).}
\label{fig:DR}
\end{figure}

The first-order equations for $T_1$, $P_1$, $n_1$, and $v_1$ are 

\begin{equation}
\dot{n_1}+\frac{i\kappa}{a}v_1=0,
\end{equation}
\begin{equation}
\dot{v_1}+\frac{\dot{a}}{a}v_1+c_{\rm s}^2\frac{i\kappa}{a}P_1=0,
\end{equation}
\begin{equation}
\dot{P_1}+\frac{\dot{a}}{a}P_1-\gamma\dot{n_1}+\frac{n_1}{\tau_{\rm net}}
+
\frac{s_{\rm T}T_1+s_{\rho}n_1}{\tau_{\rm cool}}-
\frac{r_{\rm T}T_1+r_{\rho}n_1}{\tau_{\rm heat}}
=-\ell c_{\rm s}\frac{\kappa^2}{a^2}T_1,
\end{equation}
\begin{equation}
P_1=n_1+T_1,
\end{equation}
where
\begin{equation}s_{\rm T}=\partial \ln \Lambda/\partial \ln T,\end{equation}
\begin{equation}s_{\rho} =\partial \ln \Lambda/\partial \ln n,\end{equation}
\begin{equation}r_{\rm T}=\partial \ln \Gamma/\partial \ln T,\end{equation}
\begin{equation}r_{\rho}=\partial \ln \Gamma/\partial \ln n,\end{equation}
\begin{equation}\tau_{\rm cool}^{-1}=(\gamma-1)\Lambda/k_{\rm B}T,\end{equation}
\begin{equation}\tau_{\rm heat}^{-1}=(\gamma-1)\Gamma/k_{\rm B}T.\end{equation}
We adopt the collision mean free path, $\ell$, instead of 
thermal conduction coefficient $K=nk_{\rm B}\ell c_{\rm s}/(\gamma-1)$. 
The term $\dot{a}/a$ remains constant ($-1/\gamma \tau_{\rm net}$)  
as long as $t\ll \gamma\tau_{\rm net}$.
We assume that the perturbation is proportional to $\exp(\sigma t)$
and obtain the dispersion relation as
\begin{equation}
\left\{\sigma^2-\frac{2\sigma}{\gamma\tau_{\rm net}}+(k c_{\rm s})^2\right\}
\left(\sigma
+\frac{s_{\rm T}-1}{\tau_{\rm cool}}
-\frac{r_{\rm T}-1}{\tau_{\rm heat}}+\ell c_{\rm s} k^2\right)
-(k c_{\rm s})^2\left\{(1-\gamma)\sigma+\frac{s_{\rho}}{\tau_{\rm cool}}
-\frac{r_{\rho}}{\tau_{\rm heat}}\right\}=0,
\label{eq:DR}
\end{equation} 
where $k=\kappa/a$.
This cubic equation for $\sigma$
has real root and two complex conjugate roots.  
The former root is called condensation mode, 
and the latter roots are called wave mode.
Following Field (1965), we call the most unstable mode
the condensation mode.
In Figure \ref{fig:DR},
we show an example of dispersion relation of the condensation mode.

%\clearpage

%\clearpage

\begin{deluxetable}{lrrrrrrrrrrr}
\tablecaption{MODELS. \label{tbl-1}}
\tablewidth{0pt}

\tablehead{
\colhead{Model} & 
\colhead{$n_{\rm i}$\tablenotemark{a}} & 
\colhead{$T_{\rm i}$\tablenotemark{b}} & 
\colhead{$x_{\rm ei}$\tablenotemark{c}} & 
\colhead{$N_{\rm H}$\tablenotemark{d}}  & 
\colhead{$M$\tablenotemark{e}} & 
\colhead{$V_{\rm d}$\tablenotemark{f}} & 
\colhead{$t_{\rm f}$\tablenotemark{g}} & 
\colhead{$n_{\rm f}$\tablenotemark{h}} & 
\colhead{$T_{\rm f}$\tablenotemark{i}} & 
\colhead{$P_{\rm f}/k_{\rm B}$\tablenotemark{j}} &
\colhead{$x_{\rm 2f}$\tablenotemark{k}}
} 
\startdata
W2 &0.1 &8491&  0.03 &1(20)\tablenotemark{l}
                            &2 &18 & 7.0(6)& 1.73 & 6353 & 3.3(3)&9.5(-8)\nl
W6
\tablenotemark{m} 
   &0.1 &8491&  0.03 &1(20) &6 &54 & 2.9(6)& 2.5(3)& 18.7& 5.2(4)&0.013\nl
W6L&0.1 &8912&  0.05 &1(19) &6 &54 & 2.8(5)& 2.5(3)& 19.0& 5.2(4)&5.6(-5)\nl
W12&0.1 &8491&  0.03 &1(20) &12&109&1.45(6)&1.2(4)& 16.6& 2.0(5)&0.21\nl
I12&0.1 &8491&  1.00 &1(20) &12&109&1.45(6)&1.2(4)& 16.6& 2.0(5)&0.21\nl
C2 & 10 &107 &8.7(-4)&1(20) &2 &2.1& 7.0(5)&  2.8(2)& 28.5& 8.8(3)&5.1(-5)\nl
C10
\tablenotemark{n}
   & 10 &107 &8.7(-4)&1(20) &10&10 & 1.0(5)& 1.4(4)& 16.7& 2.5(5)&0.049\nl
\enddata

\tablenotetext{a}{The initial number density of hydrogen nuclei in cm$^{-3}$.}
\tablenotetext{b}{The initial temperature in K.}
\tablenotetext{c}{The initial electron fraction.}
\tablenotetext{d}{The total column density of hydrogen nuclei in cm$^{-2}$.}
\tablenotetext{e}{The Mach number in the center of mass frame.}
\tablenotetext{f}{The velocity difference in km/s, defined by upstream velocity
 minus downstream velocity which we set to be zero.}
\tablenotetext{g}{The final time in year.}
\tablenotetext{h}{The maximum density in cm$^{-3}$ at final time step.}
\tablenotetext{i}{The temperature in K in the maximum density regions at 
 final time step.}
\tablenotetext{j}{The pressure in K cm$^{-3}$ in the maximum density
 regions at final time step.}
\tablenotetext{k}{The H$_2$ fraction in the maximum density region at
 final time step.}
\tablenotetext{l}{The notation a(b) means $a \times 10^b$}
\tablenotetext{m}{Details of this model are discussed in section \ref{subsection:WNM}}
\tablenotetext{n}{Details of this model are discussed in section \ref{subsection:CNM}}

\end{deluxetable}

%\clearpage

\begin{figure}[htbp]
\psbox[xsize=0.49\hsize]{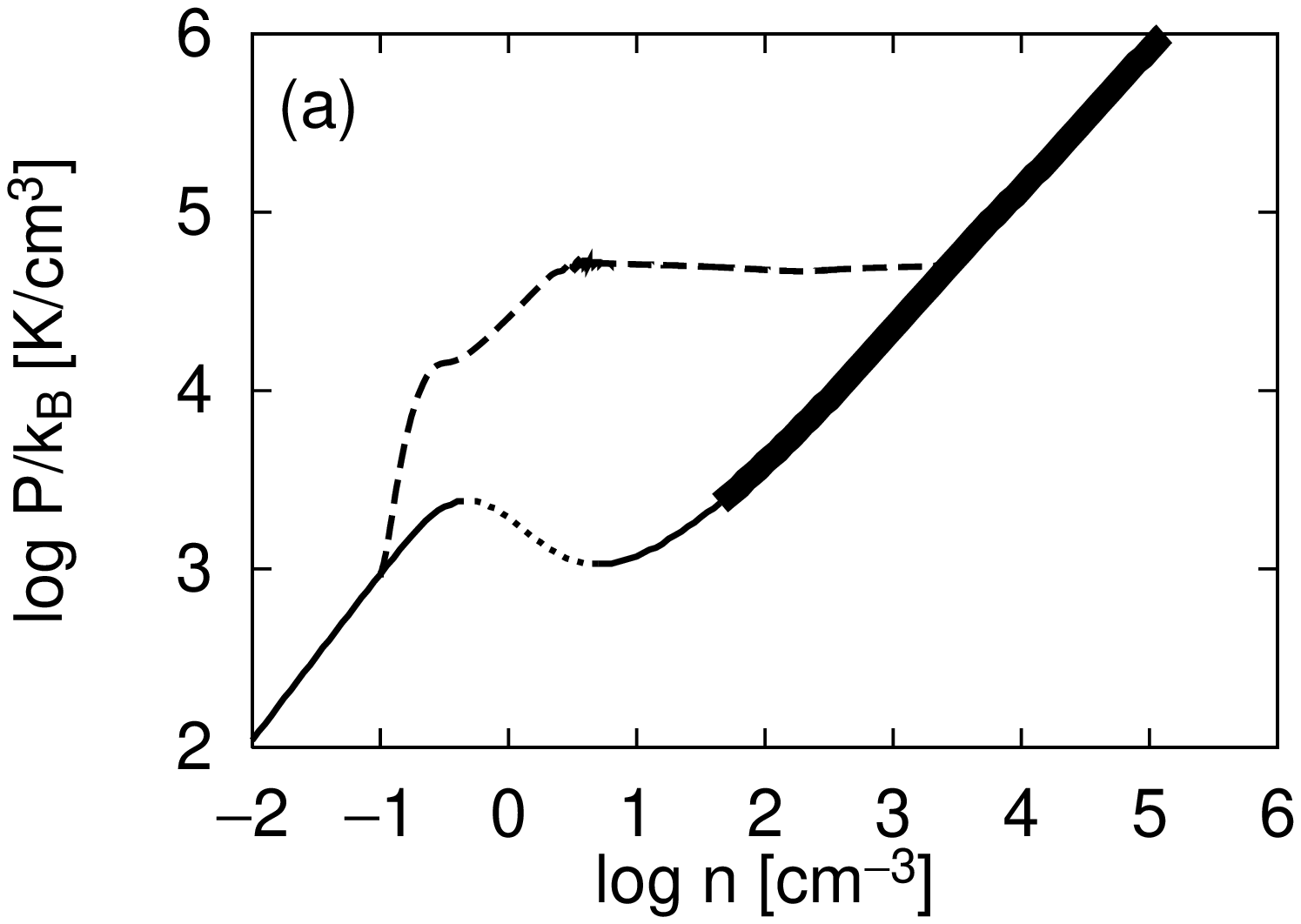}
\psbox[xsize=0.49\hsize]{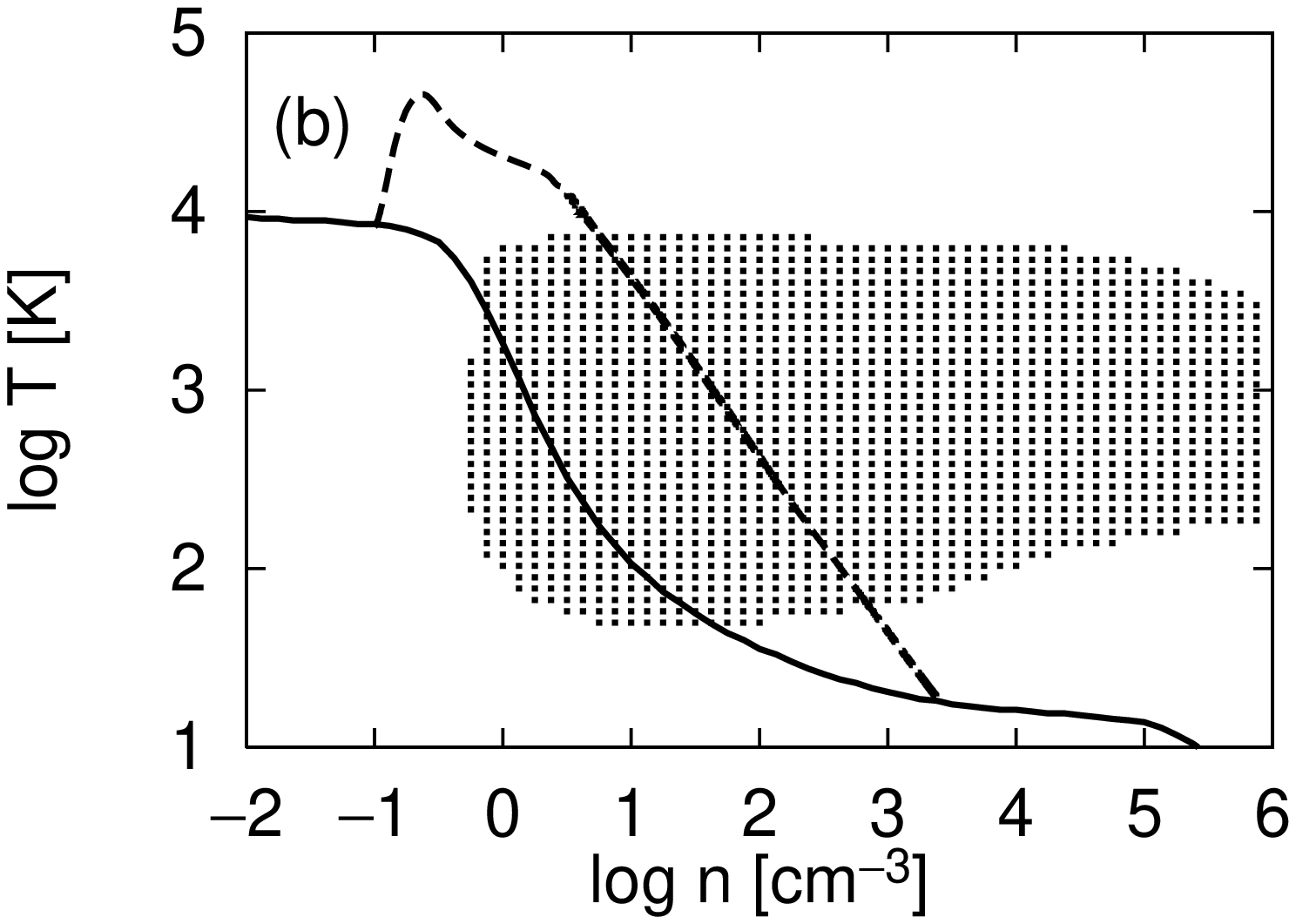}
\caption{The evolution track of WNM compression 
on the n -- P and n -- T plane.
Density, temperature and pressure evolve on the dashed lines 
from left to right.
Solid and thick lines correspond to the thermally stable equilibrium.
Dotted lines correspond to the thermally unstable equilibrium. 
In Appendix \protect\ref{LINEAR},
we investigate the stability of isobarically contracting gas
by linear perturbation theory to analyze the evolution on the dashed lines.
(b) Shaded area denotes 
thermally unstable region determined by the linear analysis.
Larger electron fraction makes cooling rate larger,
so that the unstable region is wider than the unstable region of
equilibrium gas.
This shaded region predicts that thermal instability is inevitable
when WNM cools into CNM.}
\label{fig:n-Pw,n-Tw}
\end{figure}

\begin{figure}[htbp]
\psbox[xsize=0.49\hsize]{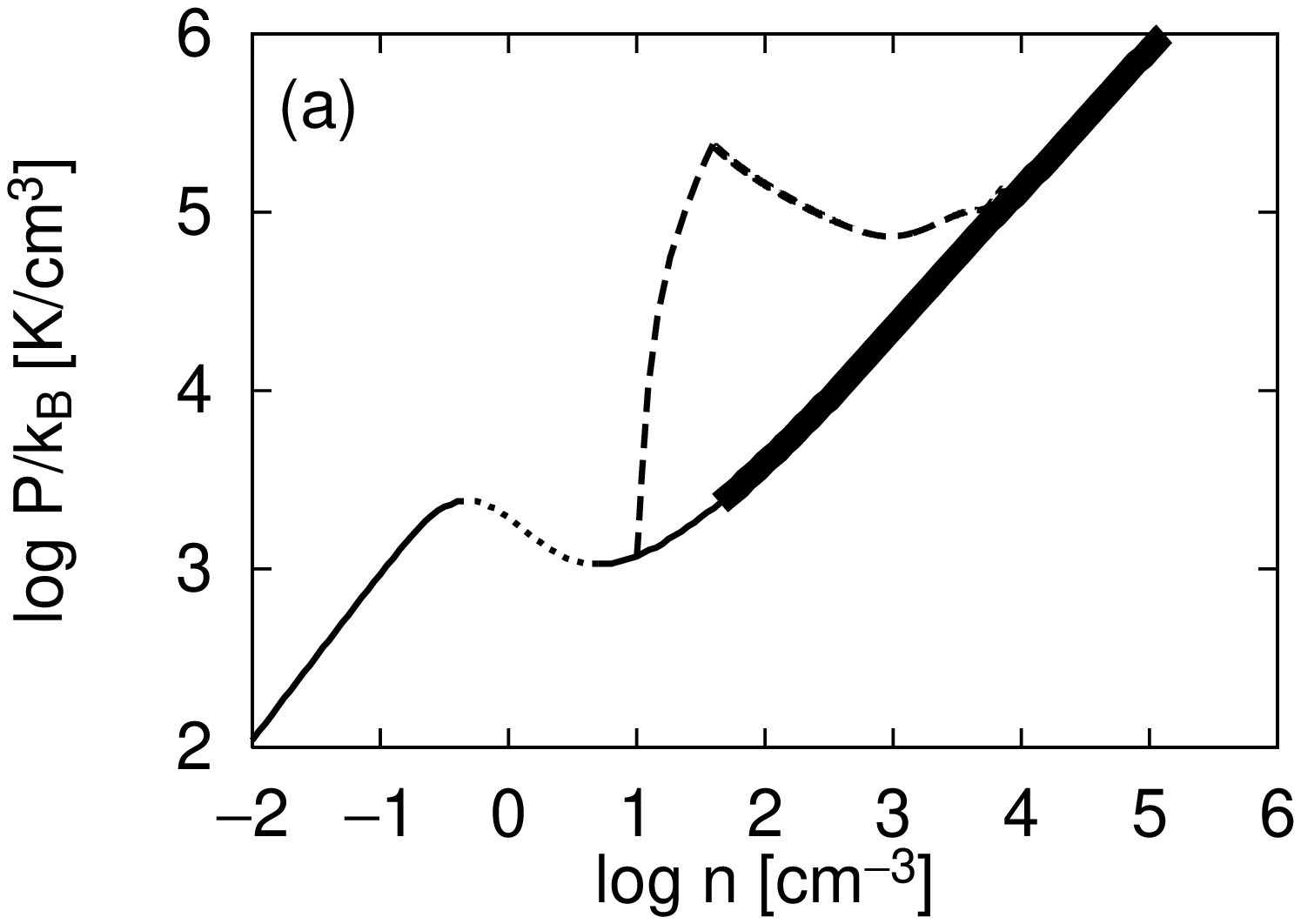}
\psbox[xsize=0.49\hsize]{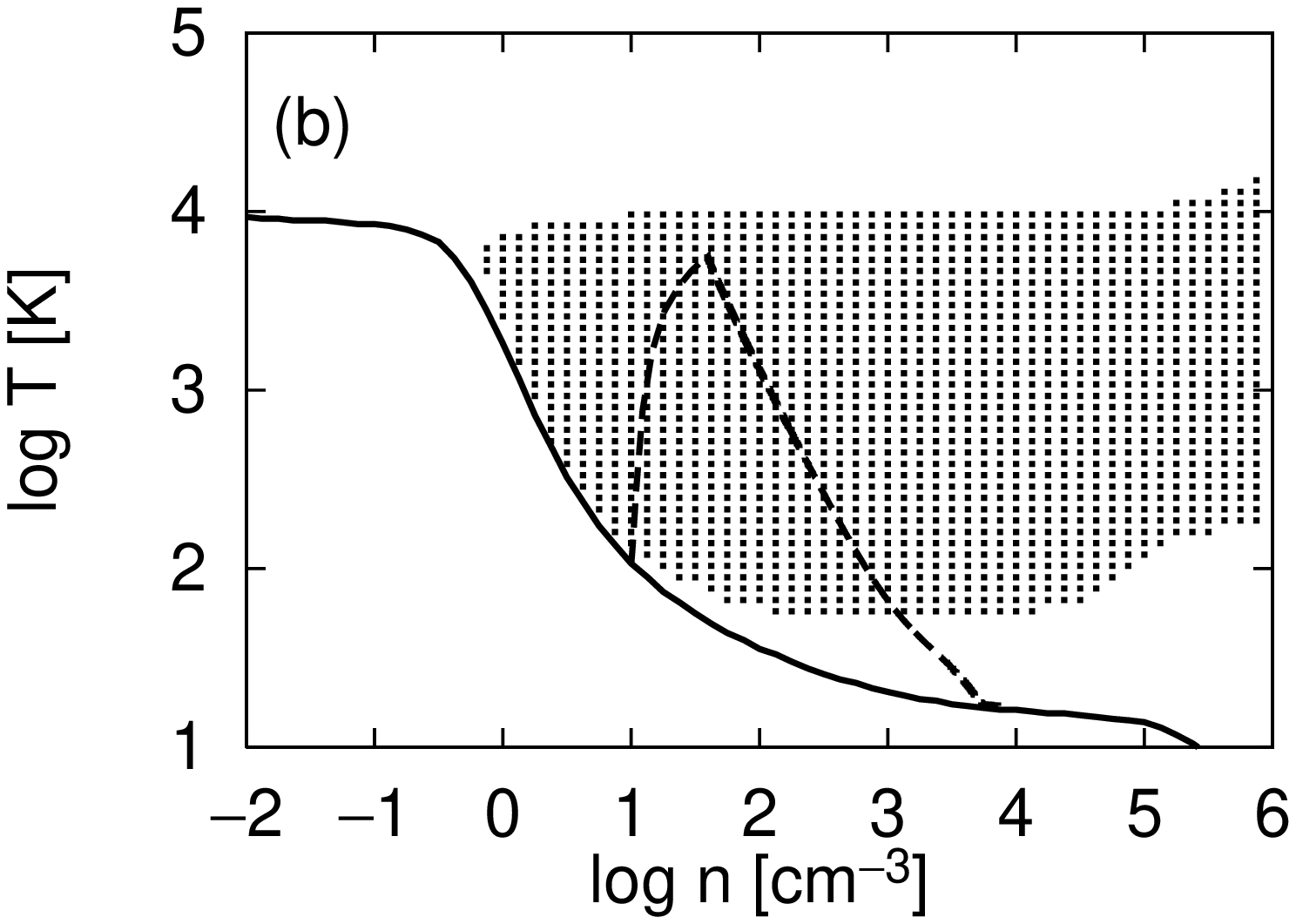}
\caption{The evolution track of CNM compression on the n -- P and n -- T plane.
Density, temperature and pressure evolve on the dashed lines 
from left to right.
Solid and thick lines correspond to the thermally stable equilibrium. 
Dotted lines correspond to the thermally unstable equilibrium. 
(b) Shaded area denotes 
thermally unstable region determined by the linear analysis.
This Figure shows how thermally stable CNM becomes unstable.}
\label{fig:n-Pc,n-Tc}
\end{figure}

\end{document}